\documentclass[10pt,journal,compsoc]{IEEEtran}
\ifCLASSOPTIONcompsoc
  \usepackage[nocompress]{cite}
\else
  \usepackage{cite}
\fi
\ifCLASSINFOpdf
\else
\fi
\usepackage{graphicx}
\usepackage{amsmath}
\usepackage{booktabs}

\usepackage{algorithm}
\usepackage{algorithmic}
\usepackage{bm}
\usepackage{amssymb}
\usepackage{subfigure}
\usepackage{multirow}
\usepackage{makecell}
\usepackage{color}
\usepackage{enumitem}
\usepackage{diagbox}
\newtheorem{example}{Motivating Example}
\newtheorem{myDef}{Definition}

\begin{document}
%
\title{A Unified Framework for Cross-Domain and Cross-System Recommendations}
%
%
%
%

\author{Feng~Zhu, 
        Yan~Wang,~\IEEEmembership{Senior~Member,~IEEE,}
        Jun~Zhou,~\IEEEmembership{Senior~Member,~IEEE,}
        Chaochao~Chen$^*$\thanks{$^*$Chaochao~Chen is the corresponding author.}, 
        Longfei~Li and
        Guanfeng~Liu
        
\IEEEcompsocitemizethanks{\IEEEcompsocthanksitem F. Zhu, J. Zhou, C. Chen, and L. Li are with Ant Group, Hangzhou 310012, China.\protect\\
E-mail: \{zhufeng.zhu, jun.zhoujun, chaochao.ccc\}@antgroup.com, longyao.llf@antgroup.com. \\
\IEEEcompsocthanksitem Y. Wang and G. Liu are with the Department of Computing, Macquarie University, Sydney, NSW 2109, Australia.\protect\\
E-mail: \{yan.wang, guanfeng.liu\}@mq.edu.au}
\thanks{}
}

%
%

\markboth{IEEE TRANSACTIONS ON KNOWLEDGE AND DATA ENGINEERING}%
{Zhu \MakeLowercase{\textit{et al.}}}
%



\IEEEtitleabstractindextext{%
\begin{abstract}
Cross-Domain Recommendation (CDR) and Cross-System Recommendation (CSR) have been proposed to improve the recommendation accuracy in a target dataset (domain/system) with the help of a source one with relatively richer information. However, most existing CDR and CSR approaches are single-target, namely, there is a single target dataset, which can only help the target dataset and thus cannot benefit the source dataset. In this paper, we focus on three new scenarios, i.e., Dual-Target CDR (DTCDR), Multi-Target CDR (MTCDR), and CDR+CSR, and aim to improve the recommendation accuracy in all datasets simultaneously for all scenarios. To do this, we propose a unified framework, called GA (based on \textbf{G}raph embedding and \textbf{A}ttention techniques), for all three scenarios. In GA, we first construct separate heterogeneous graphs to generate more representative user and item embeddings. Then, we propose an element-wise attention mechanism to effectively combine the embeddings of common entities (users/items) learned from different datasets. Moreover, to avoid negative transfer, we further propose a \textbf{P}ersonalized training strategy to minimize the embedding difference of common entities between a richer dataset and a sparser dataset, deriving three new models, i.e., GA-DTCDR-P, GA-MTCDR-P, and GA-CDR+CSR-P, for the three scenarios respectively. Extensive experiments conducted on four real-world datasets demonstrate that our proposed GA models significantly outperform the state-of-the-art approaches.
\end{abstract}

}

\maketitle

\IEEEdisplaynontitleabstractindextext

%
\IEEEpeerreviewmaketitle

\IEEEraisesectionheading{\section{Introduction}\label{sec:Introduction}}
\subsection{Background}
\begin{figure*}[t]
 	\begin{center}
 	\subfigure[\textbf{Conventional Scenario 1}: Single-target CDR]{
 			\label{fig:subfig:STCDR}
 			\includegraphics[width=0.485\textwidth]{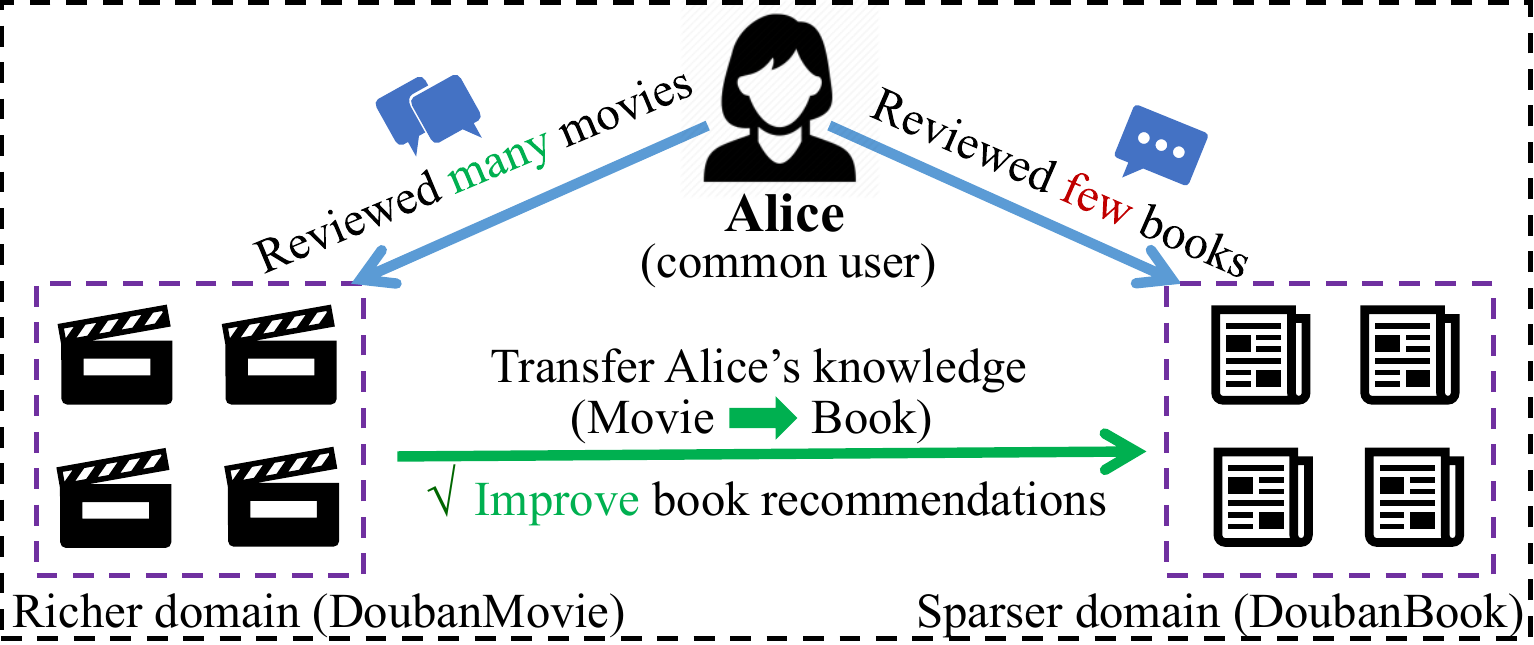}}
 	\subfigure[\textbf{Conventional Scenario 2}: Single-target CSR]{
 			\label{fig:subfig:STCSR}
 			\includegraphics[width=0.485\textwidth]{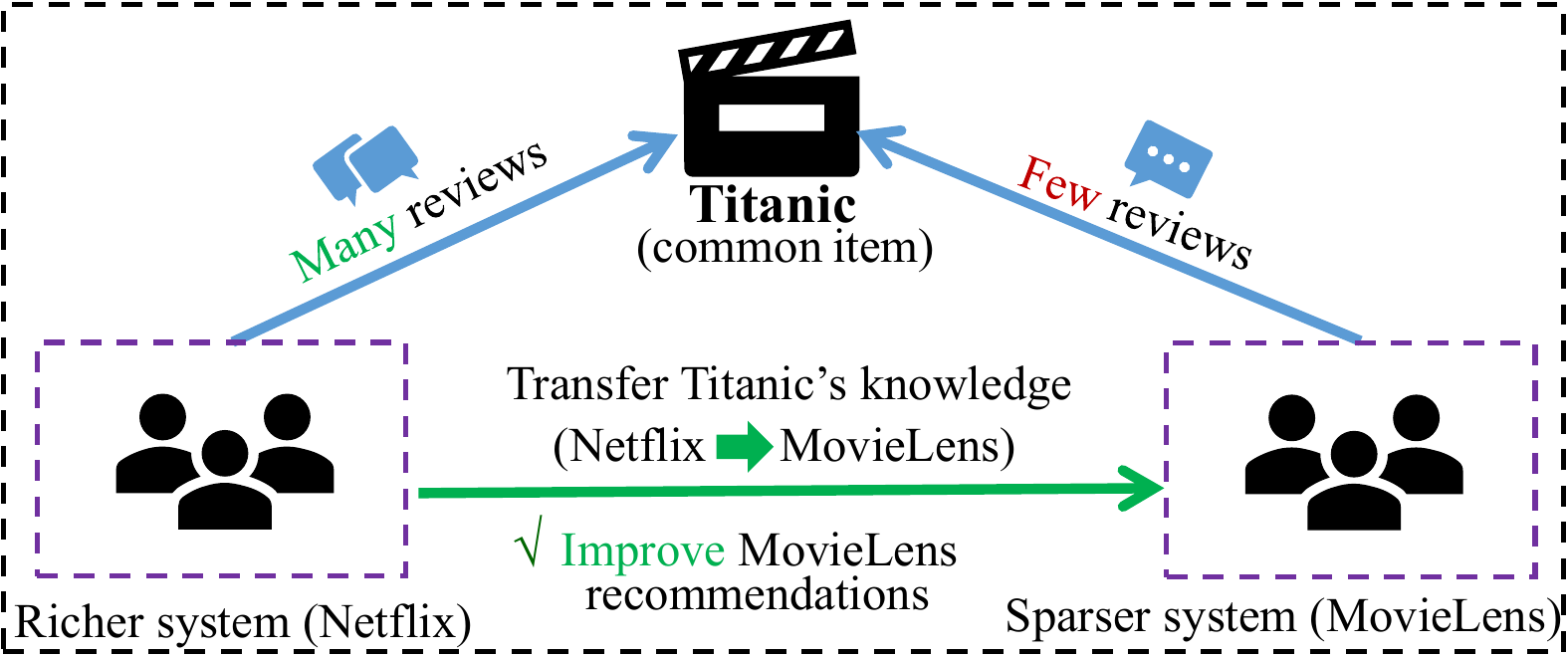}}
 	 \subfigure[The limitation of conventional single-target CDR]{
 		\label{fig:subfig:STCDR_limitation}
 		\includegraphics[width=0.485\textwidth]{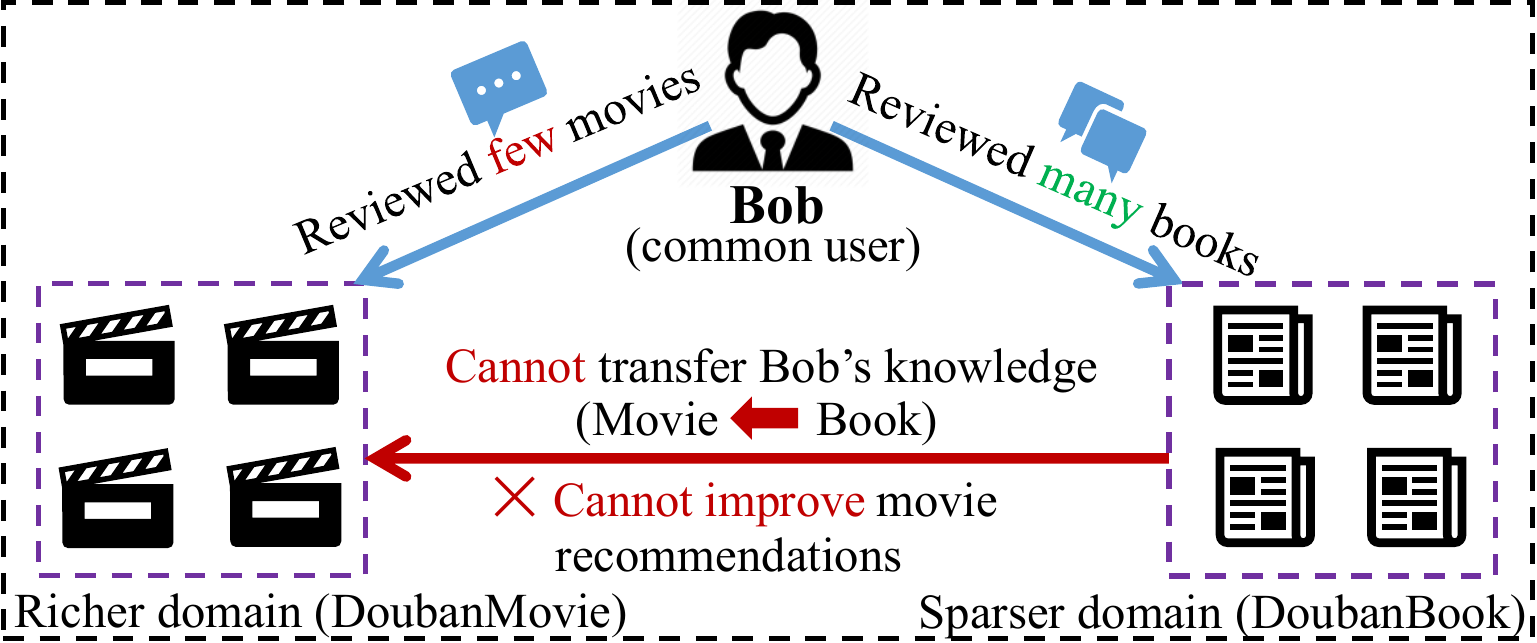}}
 	\subfigure[\textbf{Our Target Scenario 1}: Dual-target CDR]{
 		\label{fig:subfig:DTCDR}
 		\includegraphics[width=0.485\textwidth]{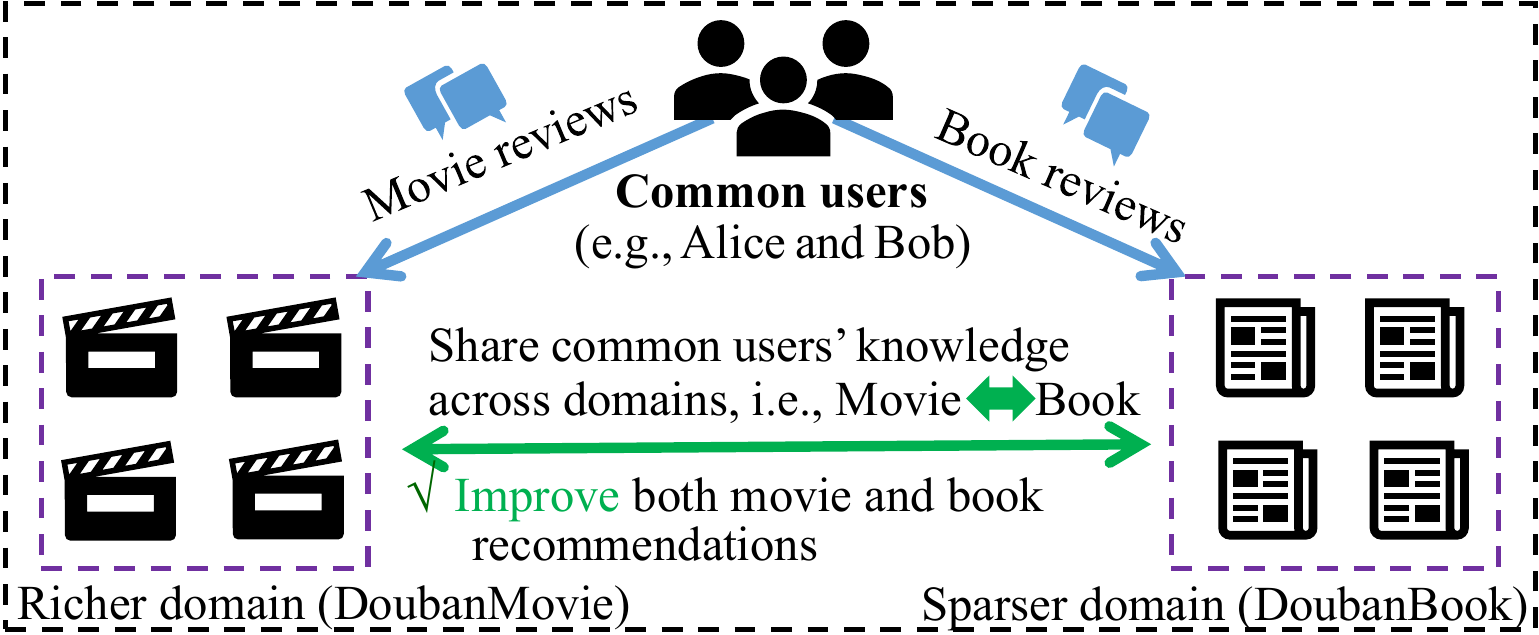}}
 	\subfigure[\textbf{Our Target Scenario 2}: Multi-target CDR]{
 		\label{fig:subfig:MTCDR}
 		\includegraphics[width=0.405\textwidth]{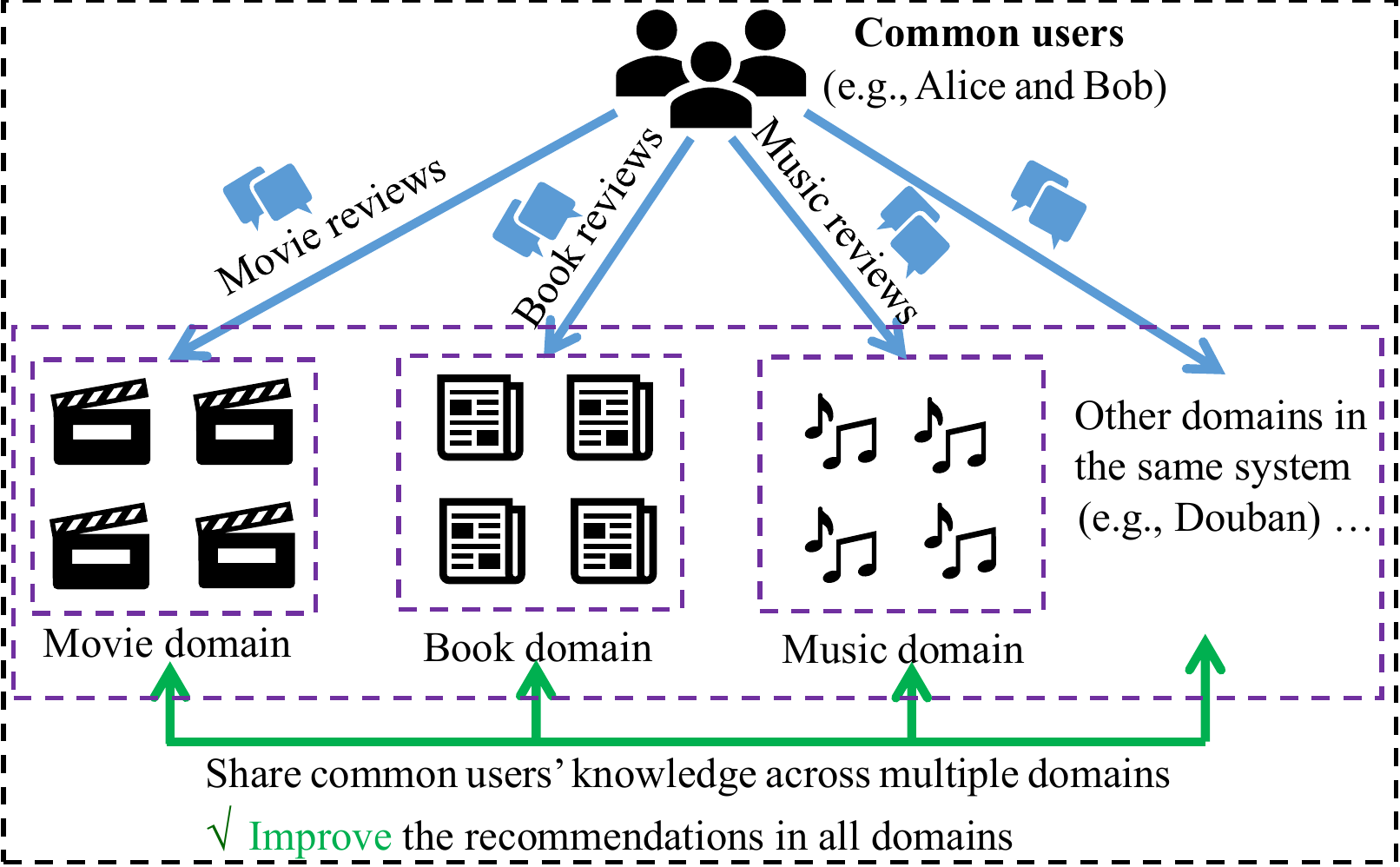}}
 	\subfigure[\textbf{Our Target Scenario 3}: CDR+CSR]{
 		\label{fig:subfig:CDR+CSR}
 		\includegraphics[width=0.575\textwidth]{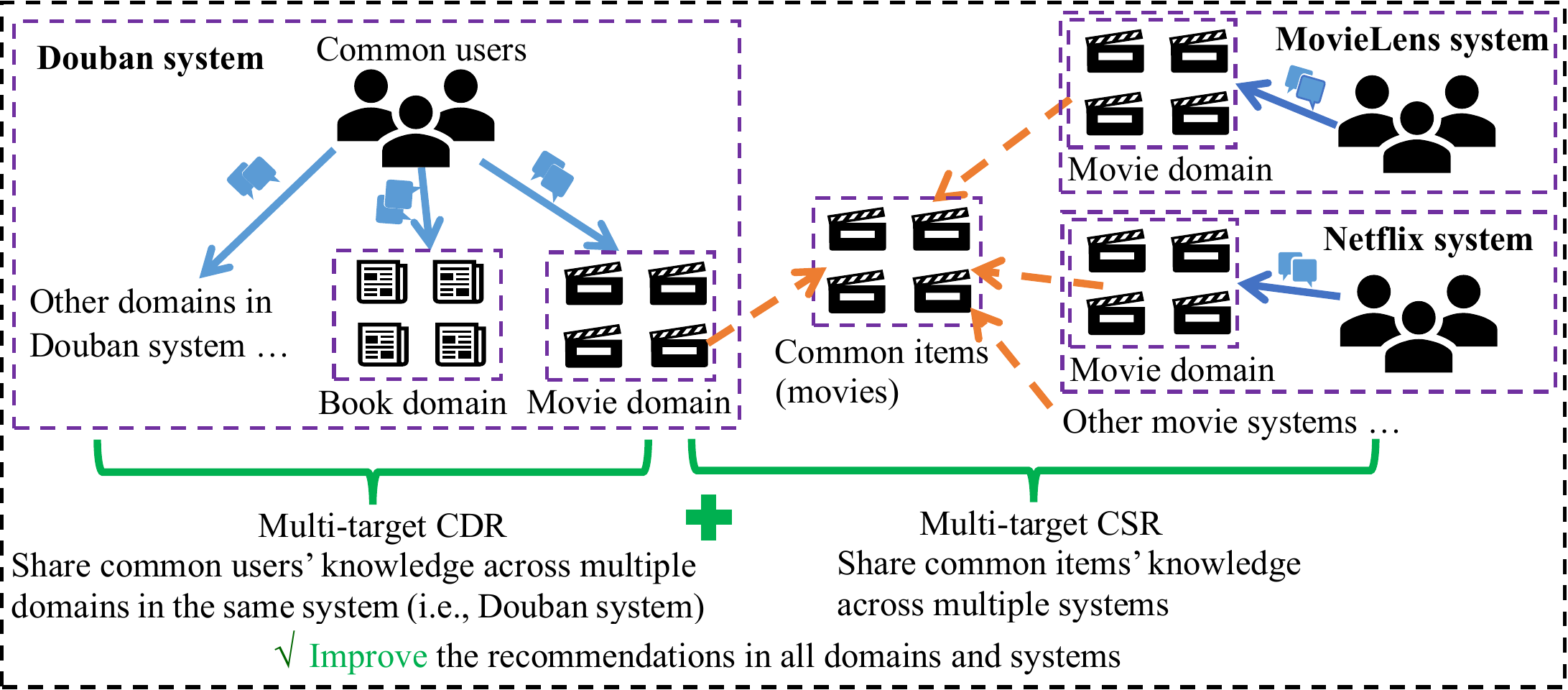}}
 	\caption{Different scenarios of cross-domain and cross-system recommendations}
 	\label{Recommendation_Categories}
 	\end{center}
 \end{figure*}

\IEEEPARstart{T}{argeting} data sparsity problem, Cross-Domain Recommendation (CDR) \cite{berkovsky2007cross} and Cross-System Recommendation (CSR) \cite{zhao2013active,zhu2018deep} have been proposed to leverage the richer information from a richer dataset (domain/system) to help improve the recommendation accuracy in a sparser one, resulting in \emph{single-target CDR} (\textbf{Conventional Scenario 1}) and \emph{single-target CSR} (\textbf{Conventional Scenario 2}). For example, in Douban system\footnote{Douban website: https://www.douban.com}, the recommender system can recommend books to a target user (e.g., Alice in Fig. \ref{fig:subfig:STCDR}) according to her movie knowledge, i.e., this is single-target CDR. In contrast, the recommender system can recommend movies to a target user in MovieLens\footnote{MovieLens website: http://www.movielens.org} according to the knowledge of these movies (e.g., Titanic in Fig. \ref{fig:subfig:STCSR}) learned from Netflix\footnote{Netflix website: https://www.netflix.com}, i.e., this is single-target CSR. In addition to the above-mentioned rating systems, CDR and CSR have been applied to other application scenarios as well, including academic searching (e.g., Arnetminer\footnote{Arnetminer website: http://arnetminer.org/} \cite{tang2012cross}), e-commerce (e.g., Amazon\footnote{Amazon website: https://www.amazon.com/} \cite{fu2019deeply}), and social networking (e.g., Facebook\footnote{Facebook website: https://www.facebook.com} \cite{shapira2013facebook} and Tencent Weibo\footnote{Tencent Weibo website: t.qq.com (it was shut down on September 28th, 2020)} \cite{jiang2012social,jiang2015social}).

CDR and CSR have different kinds of overlapping entities that serve as the `bridge' to link the two data sources. These overlapping (common) entities are relations between two domains/systems, and thus the two domains/systems are termed as \emph{related domains/systems}. In CDR, there are two related domains (e.g., movie domain and book domain) in the same system (e.g., Douban) and thus CDR techniques can utilize common users to transfer/share their knowledge across domains. Likewise, in CSR, the two related systems (e.g., Netflix and MovieLens) have the same domain (e.g., movie domain) and thus contain common items (e.g., movies). Technically, in solutions, we only need to replace the `bridge' from common users in a CDR model to common items so as to support CSR, and vice versa. This means that CDR and CSR techniques can be applied to each other's scenarios. Thus, in this paper, our proposed approaches can be applied for all related domains/systems. In this paper, without a special explanation, we basically focus on CDR when discussing solutions, except the scenario of CDR+CSR. 

\subsection{Limitations of Conventional Single-Target CDR}
Existing single-target CDR approaches can be generally classified into two groups: content-based transfer approaches and feature-based transfer approaches. \emph{Content-based transfer} tends to link different domains by identifying similar content information --- such as user profiles, item details \cite{berkovsky2007cross}, user-generated reviews \cite{tan2014cross}, and social tags \cite{fernandez2014exploiting}.  \emph{Feature-based transfer} \cite{zhang2017cross,liu2017mcs,zhu2018deep,zhang2018cross2,hu2019transfer,manotumruksa2019cross,huang2019lscd,li2019cross,zhao2020catn,kang2020deep} first trains different Collaborative Filtering (CF) based models --- such as Bayesian Personalised Ranking (BPR) \cite{rendle2009bpr}, Neural Matrix Factorization (NeuMF) \cite{he2017neural}, and Deep Matrix Factorization (DMF) \cite{xue2017deep}, to obtain user/item embeddings or patterns, and then transfers these embeddings through common or similar users across domains. In contrast to the content-based transfer approaches, feature-based transfer approaches typically employ machine learning techniques --- such as transfer learning \cite{zhang2016multi} and neural networks \cite{man2017cross}, to transfer knowledge across domains. 

\begin{example}\label{motivating_example}
	Fig. \ref{fig:subfig:STCDR_limitation} depicts a special case in the conventional single-target CDR system (i.e., Douban) that contains two domains --- DoubanMovie (the richer domain) and DoubanBook (the sparser domain) --- including users, items (movies or books), and interactions (e.g., ratings and reviews). In contrast to Alice in Fig. \ref{fig:subfig:STCDR}, who is one of majority users in the dataset, Bob in Fig. \ref{fig:subfig:STCDR_limitation}, who is one of minority users in the dataset, reviewed few movies and many books, and thus Bob's knowledge (e.g., user embedding) in the book domain would be more accurate than his knowledge in the movie domain. However, the knowledge in the book domain cannot be used to improve the knowledge in the movie domain since the conventional single-target CDR system can only leverage the information from the richer domain to improve the recommendation accuracy in the sparser domain.
\end{example}

However, all these existing single-target CDR approaches only focus on how to leverage the source domain to help improve the recommendation accuracy in the target domain, but not vice versa. This is also explained in Motivating Example \ref{motivating_example}. In fact, each of the two domains may be relatively richer in certain types of information (e.g., ratings, reviews, user profiles, item details, and tags); if such information can be leveraged well, it is likely to improve the recommendation performance in both domains simultaneously, rather than in a single target domain only. Therefore, the novel \emph{dual-target CDRs} \cite{zhu2019dtcdr,li2019ddtcdr,zhu2020graphical,li2021dual} have been recently proposed to improve the recommendation accuracy in both richer and sparser domains simultaneously by making good use of the information or knowledge from both domains.

\subsection{Our Target Scenarios}
Dual-target CDR is our first target scenario. 
Intuitively, based on the existing single-target CDR approaches, it seems to be a solution for dual-target CDR (Target Scenario 1, see Fig. \ref{fig:subfig:DTCDR}) by simply changing their transfer direction from ``Richer$\rightarrow$Sparser" to ``Sparser$\rightarrow$Richer". However, as referred to as \emph{Negative Transfer} \cite{pan2009survey}, this idea does not work, because, in principle, the knowledge learned from the sparser domain is less accurate than that learned from the richer domain, and thus, the recommendation accuracy in the richer domain is more likely to decline by simply and directly changing the transfer direction. Therefore, dual-target CDR/CSR demands novel and effective solutions.

Additionally, inspired by dual-target CDR, multi-target CDR (Target Scenario 2, see Fig. \ref{fig:subfig:MTCDR}), namely, improving the recommendation accuracy in multiple domains simultaneously, is also an interesting and challenging research problem for CDR. However, unlike dual-target CDR, in Target Scenario 2, more non-IID (independent and identically distributed) data from multiple domains may negatively affect the recommendation performance, which is likely to cause negative transfer. This is the new challenge. Though there are no solutions reported in the literature yet, multi-target CDR is similar to Multi-Domain Recommendation (MDR) to some extent. Nevertheless,  MDR \cite{zhang2012multi,moreno2012talmud,pan2013transfer,zhang2016multi} tends to improve the recommendation accuracy in a single target domain or the recommendation accuracy of a mixed user set from multiple domains by leveraging the auxiliary information from multiple domains. Therefore, a feasible multi-target CDR/CSR solution is in demand.

Moreover, it would be promising to devise a hybrid approach that can leverage the auxiliary information from both multiple domains and multiple systems to further improve the accuracy in these domains and systems simultaneously, i.e., CDR+CSR (Target Scenario 3, see Fig. \ref{fig:subfig:CDR+CSR}). This means that CDR+CSR should utilize the information of both common users and common items in the same approach. This is also an interesting and challenging research problem.

\subsection{Challenges}
Targeting Scenario 1, there are two challenges (\textbf{CH1} and \textbf{CH2}) as follows.

\textbf{CH1}: \emph{how to leverage the data richness and diversity to generate more representative single-domain user and item embeddings for improving recommendation accuracy in each of the domains?} Both traditional Collaborative Filtering (CF) models, e.g., BPR \cite{rendle2009bpr}, and novel neural CF models, e.g., NeuMF \cite{he2017neural} and DMF \cite{xue2017deep}, are based on the user-item relationship to learn user and item embeddings. However, most of them ignore the user-user and item-item relationships, and thus can hardly enhance the quality of embeddings.

\textbf{CH2}: \emph{how to effectively optimize the user or item embeddings in each target domain for improving recommendation accuracy?} The state-of-the-art dual-target CDR approaches either adopt fixed combination strategies, e.g., average-pooling, max-pooling, and concatenation \cite{zhu2019dtcdr, liu2020cross}, or simply adapt the existing single-target transfer learning to dual transfer learning \cite{li2019ddtcdr}. However, none of them can effectively combine the embeddings of common entities, and thus it is hard to achieve an effective embedding optimization in each target domain.

Targeting Scenario 2, there is a new challenge (\textbf{CH3}).

\textbf{CH3}: \emph{how to avoid negative transfer when combining the embeddings of common users from multiple domains?} Compared with dual-target CDR (Target Scenario 1), the core goal of multi-target CDR (Target Scenario 2) is to leverage more auxiliary information from more domains to improve the recommendation performance. However, it is worth noting that more non-IID data from more domains may negatively affect the recommendation performance. This is because such incomplete non-IID data, especially in sparser domains, can only reflect biased features of common users. Therefore, in Target Scenario 2, the recommendation performance in some domains may decline as more sparser domains join in, i.e., the negative transfer can thus happen.

Targeting Scenario 3, there is a new challenge (\textbf{CH4}).

\textbf{CH4}: \emph{how to effectively leverage the auxiliary information of both common users and common items simultaneously?} In a dual-target or multi-target CDR scenario, we only need to optimize the embeddings of common users from dual or multiple domains. Then, based on CF models in each domain, the embeddings of distinct users and items can be optimized gradually. However, in a CDR+CSR scenario, it should effectively leverage the embeddings of common users and common items simultaneously, which may improve the recommendation performance in each dataset (domain/system) more quickly.

\subsection{Our Approach and Contributions} 
To address the above four challenges, in this paper, we propose a unified framework for all dual-target CDR, multi-target CDR, and CDR+CSR scenarios. The characteristics and contributions of our work are summarized as follows:
\begin{itemize}[leftmargin=*] \setlength{\itemsep}{-\itemsep}
\item We propose a \textbf{G}raphical and \textbf{A}ttentional framework, called GA, for \textbf{D}ual-\textbf{T}arget \textbf{CDR} (GA-DTCDR) scenario, which can leverage the data richness and diversity (e.g., ratings, reviews, and tags) of different datasets, share the knowledge of common entities across domains;

\item To address \textbf{CH1}, we construct a heterogeneous graph, considering not only user-item relationships (based on ratings), but also user-user and item-item relationships (based on content similarities). Then, with this heterogeneous graph, we apply a graph embedding technique, i.e., Node2vec, to generate more representative single-domain user and item embeddings for accurately capturing user and item features; 

\item To address \textbf{CH2}, we propose an element-wise attention mechanism to effectively combine the embeddings of common entities learned from dual domains, which can significantly enhance the quality of user/item embeddings and thus improve the recommendation accuracy in each of both domains simultaneously. 
\end{itemize}

It is worth mentioning that this work is an extension of our preliminary work \cite{zhu2020graphical}. In this paper, we further deliver the following contributions:
\begin{itemize}[leftmargin=*] \setlength{\itemsep}{-\itemsep}
\item Different from GA-DTCDR proposed in \cite{zhu2020graphical} that only supports dual-target CDR scenario, we extend the above proposed \textbf{GA} framework and adopt a \textbf{P}ersonalized training strategy to support all \textbf{D}ual-\textbf{T}arget \textbf{CDR} (GA-DTCDR-P), \textbf{M}ulti-\textbf{T}arget \textbf{CDR} (GA-MTCDR-P), and \textbf{CDR+CSR} (GA-CDR+CSR-P) scenarios;

\item To address \textbf{CH3}, we propose a \textbf{P}ersonalized training strategy, deriving GA-DTCDR-\textbf{P} and GA-MTCDR-\textbf{P}, to train the recommendation models in different domains, which can first give personalized weights to the pair-wise embedding differences of common users between every two domains and then minimize these pair-wise embedding differences. The embeddings of common users in different domains tend to be similar but remain personalized, and thus the personalized strategy can avoid negative transfer to some extent;

\item To address \textbf{CH4}, we adjust the element-wise attention structure of GA-DTCDR to support CDR+CSR scenario and thus GA-\textbf{CDR+CSR}-P can enhance the qualities of the embeddings of common users and items simultaneously.
\end{itemize}
We conduct extensive experiments on four real-world datasets, which demonstrate that our GA-DTCDR-P significantly outperforms the best-performing baselines by an average of 9.04\% in terms of recommendation accuracy. Additionally, we conduct more multi-target CDR and CDR+CSR experiments (see Tasks 4 and 5 in \textbf{Experiments and Analysis}) to demonstrate that our GA-MTCDR-P and GA-CDR+CSR-P can further improve the best-performing baselines by an average of 9.21\%. 

\section{Related Work}
\subsection{Single-Target CDR} \label{RelatedWork-STCDR}
Most of the existing single-target CDR approaches tend to leverage auxiliary information from the source domain to improve the recommendation accuracy in the target domain. According to their transfer strategies, these single-target CDR approaches are classified into two categories: content-based transfer and feature-based transfer.

\begin{itemize}[leftmargin=*] \setlength{\itemsep}{-\itemsep}
	\item \textbf{Content-based transfer.} These approaches first link the richer and sparser domains by content information, e.g., user/item attributes \cite{berkovsky2007cross}, tags \cite{shi2011tags,wang2020tag}, social relations \cite{jiang2012social,jiang2015social,jiang2016little}, semantic properties \cite{zhang2019cross}, thumbs-up \cite{shapira2013facebook}, text information \cite{tan2014cross}, metadata \cite{sahebi2014content}, browsing or watching history \cite{kanagawa2019cross}. Then they transfer/share user preferences or item details across domains.
	
	\item \textbf{Feature-based transfer.} These approaches tend to employ some classical machine learning techniques --- such as multi-task learning \cite{lu2018like}, transfer learning \cite{hu2018conet,he2018robust,zhang2018cross2,hu2019transfer,manotumruksa2019cross,huang2019lscd,li2019cross,zhao2020catn,kang2020deep}, clustering \cite{wang2019solving}, reinforcement learning \cite{liu2018transferable}, deep neural networks \cite{man2017cross,zhu2018deep,fu2019deeply,liu2020exploiting}, relational learning \cite{sopchoke2018explainable} and semi-supervised learning \cite{kang2019semi}, to map or share features, e.g., user/item lembeddings and rating patterns \cite{he2018robust,yuan2019darec}, learned by CF-based models (e.g., classical factorization models and novel neural CF models), across domains.
\end{itemize}

Additionally, some studies \cite{zhang2012multi,pan2013transfer,zhang2016multi,zhang2020cross} focus on a derivational problem, i.e., multi-domain recommendation, which is to improve the recommendation accuracy on the target domain by leveraging the auxiliary information from multiple domains. However, all of them are single-target models, which means they cannot improve the recommendation accuracy in the richer domain even if the sparser domain may contain certain types of auxiliary information to support the richer domain.

\subsection{Dual-Target CDR}  \label{RelatedWork-DTCDR}
Dual-target CDR is still a novel concept for improving the recommendation accuracy in both domains simultaneously. Therefore, existing solutions are limited. The existing dual-target CDR approaches mainly focus on applying fixed combination strategies \cite{zhu2019dtcdr,liu2020cross}, or they focus on simply changing the existing single-target transfer learning to become dual-transfer learning \cite{li2019ddtcdr,li2021dual}. However, none of them can effectively combine the embeddings of common users.

In \cite{zhu2019dtcdr}, Zhu et al. proposed the DTCDR, which is the first dual-target CDR framework in the literature that uses multi-source information to generate more representative embeddings of users and items. Based on multi-task learning, the DTCDR framework uses three different combination strategies, e.g., average-pooling, max-pooling, and concatenation, to combine and share the embedding of common users across domains. Later on, similarly, in \cite{liu2020cross}, Liu et al. also use a fixed combination strategy, i.e., hyper-parameters and data sparsity degrees of common users.

In addition, in \cite{li2019ddtcdr}, Li et al. proposed the DDTCDR, a deep dual-transfer framework for dual-target CDR. The DDTCDR framework considers the bidirectional latent relations between users and items and applies a latent orthogonal mapping to extract user preferences. Based on the orthogonal mapping, DDTCDR can transfer users' embeddings in a bidirectional way (i.e., Richer $\rightarrow$ Sparser and Sparser $\rightarrow$ Richer). Recently, Li et al. proposed an improved version of DDTCDR in \cite{li2021dual}, i.e., a dual metric learning (DML) model for dual-target CDR.
\subsection{Graph Embedding} \label{RelatedWork-GraphEmbedding}

Graph Embedding is to learn a mapping function that maps the nodes in a graph to low-dimensional latent representations \cite{zhou2018graph}. These latent representations can be used as the features of nodes for different tasks, such as classification and link prediction. According to embedding techniques, this section classifies the existing graph-embedding approaches into two categories: dimensionality reduction and neural networks. Dimensionality reduction-based approaches --- such as multidimensional scaling \cite{kruskal1978multidimensional}, principal component analysis \cite{wold1987principal} and their extensions \cite{yan2005graph} --- involve optimising a linear or non-linear function that reduces the dimension of a graph's representative data matrix and then produces low-dimensional embeddings. Neural network-based approaches --- such as DeepWalk \cite{perozzi2014deepwalk}, LINE \cite{tang2015line} and Node2vec \cite{grover2016node2vec} --- involve treating nodes as words and the generated random walks on graphs as sentences, and then learning node embeddings based on these words and sentences \cite{zhou2018graph}. Also, recently, there are some graph embedding approaches that can leverage both explicit preferences and heterogeneous relationships by graph convolutional networks \cite{zhao2019intentgc,yang2020interpretable}.

\subsection{Attention Mechanism} \label{RelatedWork-Attention}
Attention is firstly introduced in \cite{bahdanau2014neural}, which provides more accurate alignment for each position in a machine translation task. Apart from machine translation, recently, attention mechanism also has been widely used in recommendation \cite{chen2017attentive}. The general idea of the attention mechanism is to focus on selective parts of the whole information, which can capture the outstanding features of objects. For recommendation, the existing attention approaches \cite{hu2018leveraging,tay2018multi,wang2019kgat} tend to select more informative parts of explicit or implicit data to improve the representations for users and items.

\begin{table}
		\caption{Important notations} \label{Notations}
		\begin{center}
				\resizebox{\textwidth/2}{!}{
			\begin{tabular}{|c|c|}
				\hline
				\textbf{Symbol} & \textbf{Definition}\\
				\hline
				$c_{ij} \in C$ & \makecell{the comment (e.g., the review and the tags) of \\user $u_i$ on item $v_j$}\\
				\hline
				$C \in \mathbb{R}^{m \times n}$ & the user comments\\
				\hline
				$CD=\{cd_1,cd_2,...,cd_{m+n}\}$ & \makecell{the content documents of users and items}\\
				\hline
				$D^x$ & Domain $x$\\
				\hline
				$ID=\{id_1,...,id_n\}$ & the item details\\
				\hline
				$G=(\{\mathcal{U},\mathcal{V}\}, E)$ & \makecell{the heterogeneous graph, $E$ is the set of \\user-user, user-item, and item-item relationships} \\
				\hline
				$k$ & the dimension of embedding matrix\\
				\hline
				$m$ & the number of users\\
				\hline
				$n$ & the number of items\\
				\hline
				$\tilde{U}$ & the combined embeddings of common users\\
				\hline
				$r_{ij} \in R $ & the rating of user $u_i$ on item $v_j$\\
				\hline
				$R \in \mathbb{R}^{m \times n}$ & the rating matrix\\
				\hline
					$S^x$ & System $x$\\
				\hline
				$\mathcal{U}=\{u_1,...,u_m\}$   & the set of users \\
				\hline
				$U$ & the graph embedding matrix of users\\
				\hline
				$UC$ & the document embedding matrix of users\\
				\hline
				$UP=\{up_1,...,up_m\}$ & the user profiles\\
				\hline
				$\mathcal{V}=\{v_1,...,v_n\}$ &the set of items\\
				\hline
				$V$ & the graph embedding matrix of items\\
				\hline
				$VC$ & the document embedding matrix of items\\
				\hline
				$y_{ij} \in Y$ & the interaction of user $u_i$ on item $v_j$\\
				\hline
				$Y \in \mathbb{R}^{m \times n}$ & the user-item interaction matrix\\
				\hline
				$*^{x}, x \in \{1,2,...,a\}$ & \makecell{the notations for domain $x$, where $a$\\ is the total number of domains, e.g., $m^1$ \\represents the number of users in domain $1$}\\
				\hline
				$\hat{*}$ & \makecell{the predicted notations, e.g., $\hat{y}_{ij}$ represents the \\predicted interaction of  $u_i$ on item $v_j$ }\\
				\hline
			\end{tabular}
		}
		\end{center}
\end{table}
\section{The Proposed Model} \label{ProposedModel}
In this section, we first formalize the dual-target CDR, multi-target CDR, and CDR+CSR problems. Then, we preliminarily propose a \textbf{G}raphical and \textbf{A}ttentional framework, called GA, for \textbf{DTCDR} (GA-DTCDR) scenario. Next, we extend the above \textbf{GA} framework and adopt a \textbf{P}ersonalized training strategy to support all \textbf{dual-target CDR} (GA-DTCDR-P), \textbf{multi-target CDR} (GA-MTCDR-P), and \textbf{CDR+CSR} (GA-CDR+CSR-P) scenarios. Finally, we present the detailed components of GA-DTCDR (or GA-DTCDR-P), GA-MTCDR-P, and GA-CDR+CSR-P. 
\subsection{Problem Statement} \label{ProblemStatement}
First, for the sake of better readability, we list the important notations of this paper in Table \ref{Notations}. Then, we define the Dual-Target CDR, Multi-Target CDR, and CDR+CSR as follows.

 \begin{myDef} \label{Definition_DTCDR}
 	\textbf{Dual-Target Cross-Domain Recommendation (DTCDR)}: \emph{Given two related domains $1$ and $2$, with explicit feedback (e.g., ratings and comments), implicit feedback (e.g., purchase and browsing histories), and side information (e.g., user profiles and item details), DTCDR is to improve the recommendation accuracy in both domains simultaneously by leveraging their observed information.}
\end{myDef}

\begin{myDef} \label{Definition_MTCDR}
\textbf{Multi-Target Cross-Domain Recommendation (MTCDR)}: \emph{Given multiple related domains $1$ to $a$, with explicit feedback, implicit feedback, and side information, MTCDR is to improve the recommendation accuracy in all domains simultaneously by leveraging their observed information.}
\end{myDef}

\begin{myDef} \label{Definition_CDR+CSR}
\textbf{Cross-Domain and Cross-System Recommendation (CDR+CSR)}:  \emph{Given multiple related domains/sytems $1$ to $a$, with explicit feedback, implicit feedback, and side information, CDR+CSR is to improve the recommendation accuracy in all domains and systems simultaneously by leveraging their observed information.}
\end{myDef}

Note that a certain degree of overlap between the users of different domains, i.e., common users, and overlap between the items of different systems, i.e., common items, play a key role in bridging the different datasets (domains/systems) and exchanging knowledge across them. This is a common idea of the existing CDR and CSR approaches \cite{man2017cross,zhao2017unified,zhu2018deep}.
\begin{figure}[t]
	\begin{center}
		\includegraphics[width=0.49\textwidth]{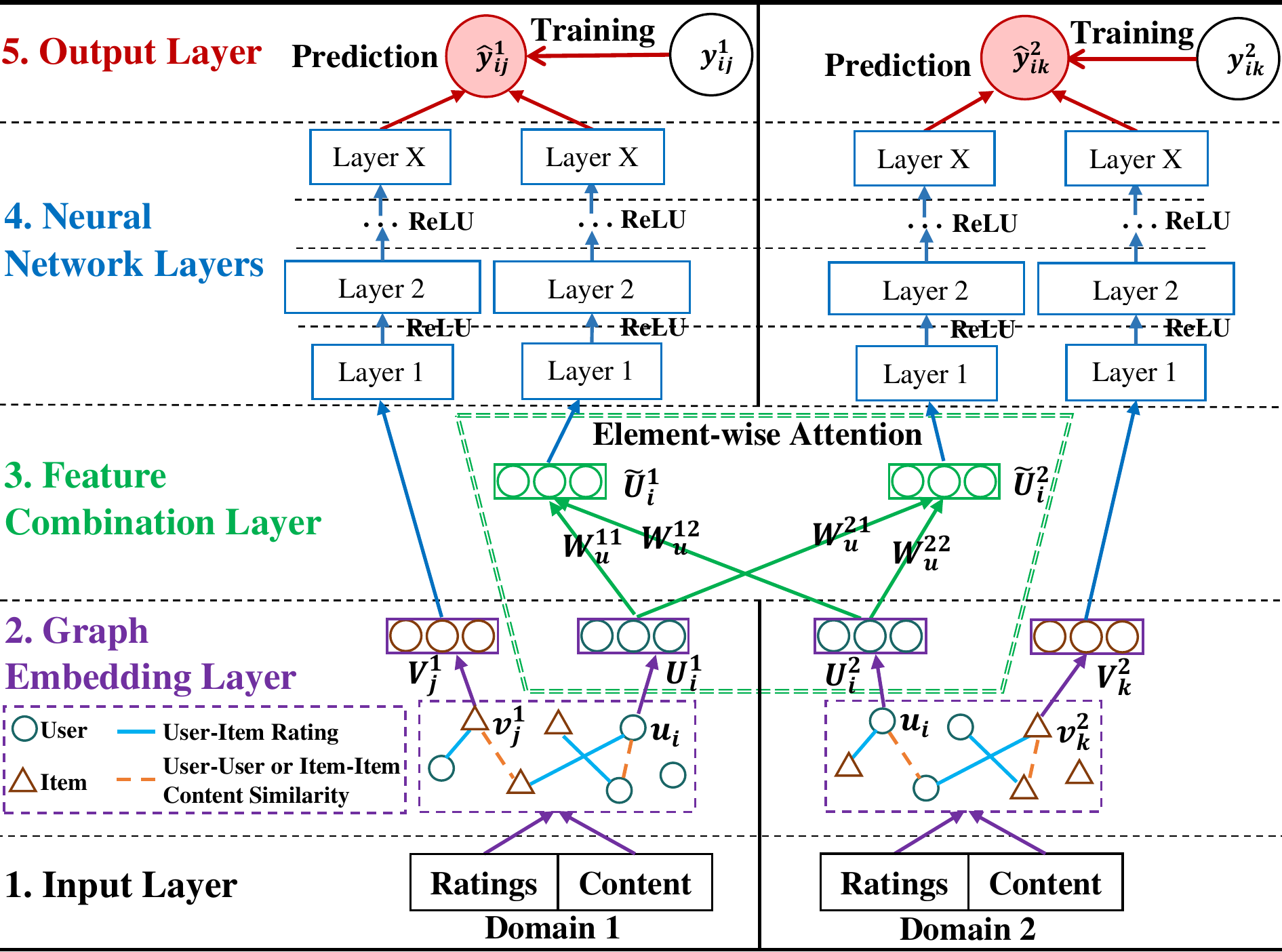}
		\caption{The overview of GA-DTCDR-P (or GA-DTCDR). This is our GA framework for DTCDR scenario, and the only difference between GA-DTCDR and GA-DTCDR-P is about their training strategies (i.e., GA-DTCDR adopts the objective function in Section \ref{Preliminary_training}, while GA-DTCDR-P adopts the personalized objective function in Section \ref{Personalized_training}). Note that, for domain $x$, $y \in\{1,2\}$, $\sum_{y=1}^{2} W^{xy}_u= 1$, where $W^{xy}_u$ is the weight vector for the embedding of common users}
		\label{GeneralFramework_DTCDR}
	\end{center}
\end{figure}

\subsection{Overview of GA Framework}\label{Section-GeneralFramework}
In this section, we first take GA-DTCDR-P (or GA-DTCDR) as an example to introduce the general structure of GA. As shown in Fig. \ref{GeneralFramework_DTCDR}, GA-DTCDR-P framework is divided into five main components, i.e., \emph{Input Layer}, \emph{Graph Embedding Layer}, \emph{Feature Combination Layer}, \emph{Neural Network Layers}, and \emph{Output Layer}. The main differences between GA-DTCDR-P and other two sub-frameworks, i.e., GA-MTCDR-P and GA-CDR+CSR-P, are the network structures of element-wise attention (see the \emph{Graph Embedding Layers} and \emph{Feature Combination Layers} of Figs. \ref{GeneralFramework_DTCDR}, \ref{GeneralFramework_MTCDR}, and \ref{GeneralFramework_CDR+CSR}). For clarity, we ignore the same components of GA-MTCDR-P and GA-CDR+CSR-P with GA-DTCDR, i.e., (1) \emph{Input Layer}, (4) \emph{Neural Network Layers}, and (5) \emph{Output Layer}. We will present the details of each component in the following sections.

Like the single-target or dual-target CDR approaches in \cite{zhao2017unified,zhu2018deep,zhu2019dtcdr}, our GA-DTCDR-P and GA-MTCDR-P can be applied to dual-target CSR and multi-target CSR as well, where the two/multiple systems have the same domain but different users, and thus contain common items only --- such as DoubanMovie and MovieLens (see Task 3 in \textbf{Experiments and Analysis}). Accordingly, in Figs. \ref{GeneralFramework_DTCDR} and \ref{GeneralFramework_MTCDR}, we only need to replace common users with common items for supporting dual-target CSR and multi-target CSR.

In fact, GA-MTCDR-P is an extension of GA-DTCDR-P from dual domains to multiple domains. GA-CDR+CSR-P is the full version of GA to handle almost all CDR and/or CSR scenarios. If there are only common users among all datasets in GA-CDR+CSR-P, GA-CDR+CSR-P will be degraded to GA-DTCDR-P or GA-MTCDR-P. Similarly, if there are only common items among all datasets in GA-CDR+CSR-P, then GA-CDR+CSR-P will be degraded for dual-target CSR or multi-target CSR.

The time complexities of GA-DTCDR-P and GA-MTCDR-P are both $\mathcal{O}(\sum_{D^x \in \mathcal{D}} i^x *(\sum_{D^x \in \mathcal{D}}m^x +k^l))$, where $i^x$ is the number of interactions in domain $D^x$, $m^x$ is the number of users in domain $D^x$ (note that for DTCSR or MTCSR, the number of users $m^x$ is replaced by the number of the number of items $n^x$ in the time complexity expression), $k$ is the number of nodes in each MLP layer (the node number is relative to the embedding dimension $k$), and $l$ is depth of MLP layers. Similarly, the time complexity of GA-CDR+CSR-P is $\mathcal{O}(\sum_{D^x \in \mathcal{D}} i^x *(\sum_{D^x \in \mathcal{D}}(m^x + n^x) +k^l))$. Compared with GA-DTCDR-P and GA-MTCDR-P, GA-CDR+CSR-P can share the embeddings of both common users and common items across domains/systems, and thus there is the sum of the number of users and the number of items, i.e., $(m^x + n^x)$, in the time complexity expression. Although $k$ and $l$ are constants in our experiments, $k^l$ is still very large. However, a deep MLP structure can represent a complex and well-trained non-linear relation between users and items, and thus can enhance the recommendation accuracy. This is a trade-off between running time and recommendation accuracy.

We now briefly present each component of GA as follows. 
\begin{itemize}[leftmargin=*] \setlength{\itemsep}{-\itemsep}
\item{\textbf{Input Layer.}} First, for the input of our GA-DTCDR-P, GA-MTCDR-P, and GA-CDR+CSR-P, we consider both explicit feedback (ratings and comments) and side information (user profiles and item details). These input data can be generally classified into two categories, i.e., rating information and content information.

\item{\textbf{Graph Embedding Layer.}} Then, we leverage rating and content information of each domain to construct a heterogeneous graph, representing user-item interaction relationships, user-user similarity relationships, and item-item similarity relationships. Based on the graph, we apply the Graph Embedding model, i.e., Node2vec \cite{grover2016node2vec}, to generate user and item embedding matrices.

\item{\textbf{Feature Combination Layer.}} Next, we propose an element-wise attention mechanism to combine the common users' embeddings from dual (GA-DTCDR-P) or multiple (GA-MTCDR-P) domains. This layer intelligently gives a set of weights to the embeddings of a common user learned from dual/multiple domains and generates a combined embedding for the common user, which remains his/her features learned from different domains with different proportions. Additionally, for GA-CDR+CSR-P, the element-wise attention mechanism can be applied to combine both the common users' embeddings and the common items' embeddings.

\item{\textbf{Neural Network Layers.}} In this component, we apply a fully-connected neural network, i.e., Multi-Layer Perceptrons (MLP), to represent a non-linear relationship between users and items in each domain.

\item{\textbf{Output Layer.}} Finally, we can generate final user-item interaction predictions. The training of our model is mainly based on the loss between predicted user-item interactions and observed user-item interactions.
\end{itemize}

Next, we will introduce the details of Graph Embedding Layer, Feature Combination Layer, Neural Network Layers, and Output Layer in the following sections.
\subsection{Graph Embedding Layer}\label{Section-GraphEmbedding}
The existing embedding strategies for recommender systems mainly focus on representing the user-item interaction relationship. Apart from the user-item interaction relationship, we use a graph to represent user-user and item-item relationships as well. Therefore, based on the rating and content information observed from dual or multiple domains, we construct a heterogeneous graph, including nodes (users and items) and weighted edges (ratings and content similarities), for each domain. Then, we can generate more representative user and item embedding matrices. The Graph Embedding contains three main sub-components, i.e., \emph{Document Embedding}, \emph{Graph Construction}, and \emph{Output}.

\subsubsection{Document Embedding} To construct the heterogeneous graph, we need to compute the content similarities between two users or two items. To this end, we consider multi-source content information, e.g., reviews, tags, user profiles, item details, observed from dual/multiple domains, to generate user and item content embedding matrices. In this paper, we adopt the most widely used model, i.e., Doc2vec \cite{le2014distributed}, as the document embedding technique. The detailed document embedding process works as follows: (1) First, in the training set, for a user $u_i$, we collect the comments (reviews and tags) $C_{i*}$ and the user profile $up_i$ of $u_i$ into the same content document $cd_i$, while for an item $v_j$, we collect the comments (reviews and tags) $C_{*j}$ on the item and its item detail $id_j$ into the same content document $cd_{m+j}$; (2) Next, we segment the words in the documents $CD=\{cd_1,cd_2,...,cd_{m+n}\}$ by using the most widely used natural language tool, i.e., StanfordCoreNLP \cite{manning2014}; (3) Finally, we apply Doc2vec model to map the documents $CD$ into the text vectors $UC$ and $VC$ for users and items, respectively.

\subsubsection{Graph Construction} First, we link the users and items via their interaction relationships. The weights of these interaction edges are normalized ratings, i.e., $R/max(R)$. To consider the user-user and item-item relationships in the heterogeneous graph, we generate the synthetic edges between two users or two items according to their normalized content similarities (edge weights). The generation probability $P(i,l)$ of the edge between users $u_i$ and $u_l$ is as follows:
\begin{equation}\label{GenerationProbability}
\begin{aligned}
&P(i,l) =\alpha \cdot sim(UC_i,UC_l),
\end{aligned}
\end{equation}
where $\alpha$ is a hyper-parameter which controls the sampling probability and $sim(UC_i,UC_l)$ is the normalized cosine similarity between $UC_i$ and $UC_l$. Similarly, we can obtain the generation probability between two items. Based on the user-item interaction relationships, user-user similarity relationships, and item-item similarity relationships, we can construct the heterogeneous graphs $G^x$ for domain $x$, where $x \in \{1,2,...,a\}$.

Similar to the approaches proposed in \cite{jiang2012social,jiang2015social}, we also construct a heterogeneous graph to represent the relations among users and items. But we construct a heterogeneous graph in each domain rather than a common graph as in \cite{jiang2012social,jiang2015social}.
\subsubsection{Output} Based on the heterogeneous graph $G^x$, we employ the graph embedding model, i.e., Node2vec \cite{grover2016node2vec}, to generate user embedding matrix $U$ and item embedding matrix $V$ for domain $x$.
\begin{figure}[t]
	\begin{center}
		\includegraphics[width=0.49\textwidth]{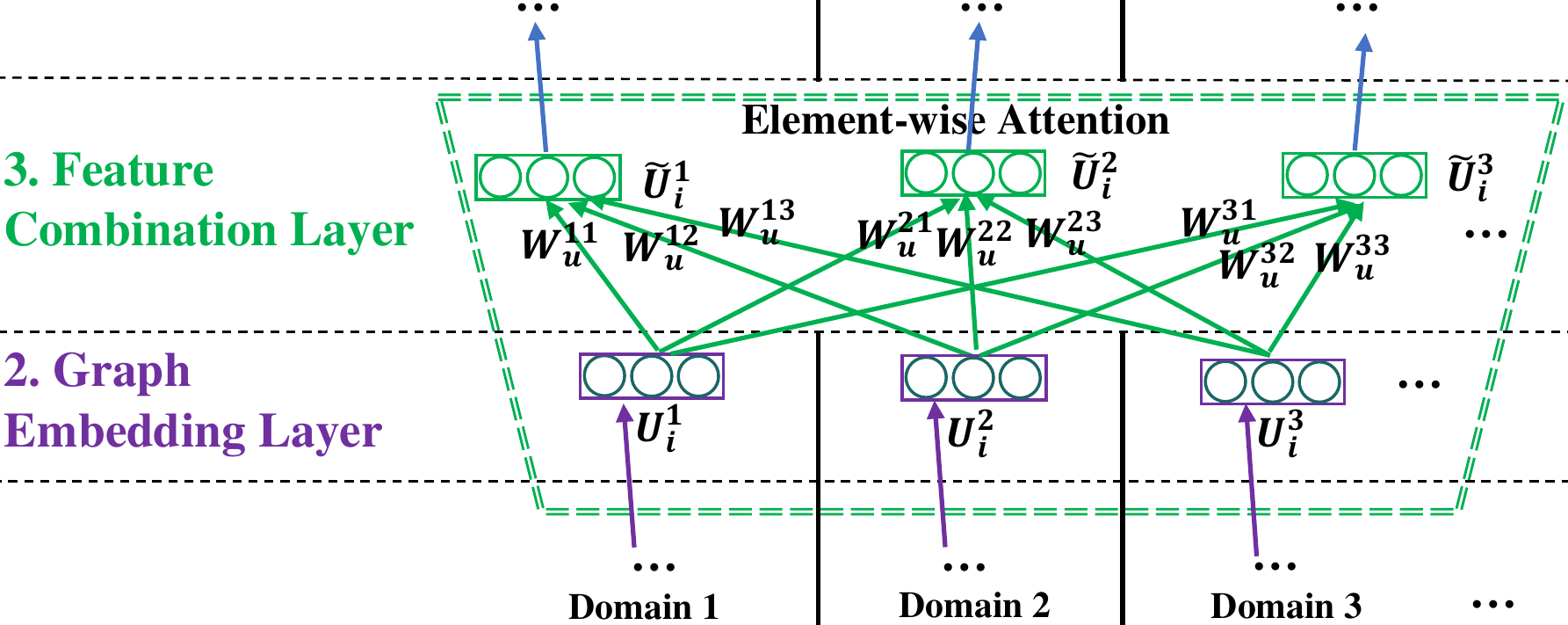}
		\caption{The overview of GA-MTCDR-P. Compared with GA-DTCDR-P, this is the extended structure for MTCDR scenario. In fact, this is an extension of GA-DTCDR-P from dual domains to multiple domains. For clarity, we ignore the same components with GA-DTCDR-P, i.e., \emph{Input Layer}, \emph{Neural Network Layers}, and \emph{Output Layer}.  Note that, for domain $x$, $x \in\{1,2,...,a\}$, $\sum_{y=1}^{a} W^{xy}_u = 1$, where $a$ is the total number of domains and $W^{xy}_u$ is the weight vector for the embedding of common users}
		\label{GeneralFramework_MTCDR}
	\end{center}
\end{figure}
\subsection{Feature Combination Layer}
Feature Combination Layer is to combine the embeddings of common entities learned from dual/multiple datasets. By doing so, the combined embeddings of common entity for each dataset can remain all features learned from the two/multiple datasets in different proportions. To this end, we propose an element-wise attention mechanism. The traditional attention mechanism tends to select a certain part of representative features and give these features higher weights when generating the combined features \cite{bahdanau2014neural}. Similarly, for a common entity, our element-wise attention mechanism tends to pay more attention to the more informative elements from each set of embedding elements (the embeddings of this common entity learned from different datasets). Compared with DTCDR and MTCDR scenarios (only common users), in CDR+CSR scenario, our element-wise attention mechanism needs to combine the embeddings of common users and items simultaneously. Therefore, we will separately introduce the feature combination layers of GA-DTCDR-P and GA-MTCDR-P and the feature combination layer of GA-CDR+CSR-P.

\subsubsection{For GA-DTCDR-P and GA-MTCDR-P} 
In GA-DTCDR-P and GA-MTCDR-P (see Figs. \ref{GeneralFramework_DTCDR} and \ref{GeneralFramework_MTCDR}), the feature combination layers are to combine the embeddings of common users learned from dual/multiple domains by our element-wise attention mechanism. For a common user $u_i$, our element-wise attention mechanism tends to pay more attention to the more informative elements from each set of elements in $\{U_i^1,U_i^2,...,U_i^a\}$, where $a$ is the total number of domains (for DTCDR, $a=2$ and for MTCDR, $a>2$). Thus our element-wise attention mechanism can generate more representative embeddings $\{\tilde{U}_i^1,\tilde{U}_i^2,...,\tilde{U}_i^a\}$ of the common user $u_i$ for domains $1, 2,..., a$, respectively. The structures of element-wise attention are shown in \emph{Feature Combination Layer} of Figs. \ref{GeneralFramework_DTCDR} and \ref{GeneralFramework_MTCDR}, respectively. The combined embedding $\tilde{U}^x_i$ of a common user $u_i$ for domain $x$ can be represented as:
\begin{equation}\label{Element-WiseAttention}
\begin{aligned}
&\tilde{U}^x_i=\sum_{y=1}^{a} W^{xy}_u \odot U^y_i,~~~\sum_{y=1}^{a} W^{xy}_u =1,
\end{aligned}
\end{equation}
where $\odot$ is the element-wise multiplication and $W^{xy}_u$ is the weight vector of the embedding of common users from domain $y$ for domain $x$.

Note that for the distinct users and all the items in each domain, we just reserve their embeddings without using the attention mechanism because they do not have dual/multiple embeddings.
\begin{figure}[t]
	\begin{center}
		\includegraphics[width=0.49\textwidth]{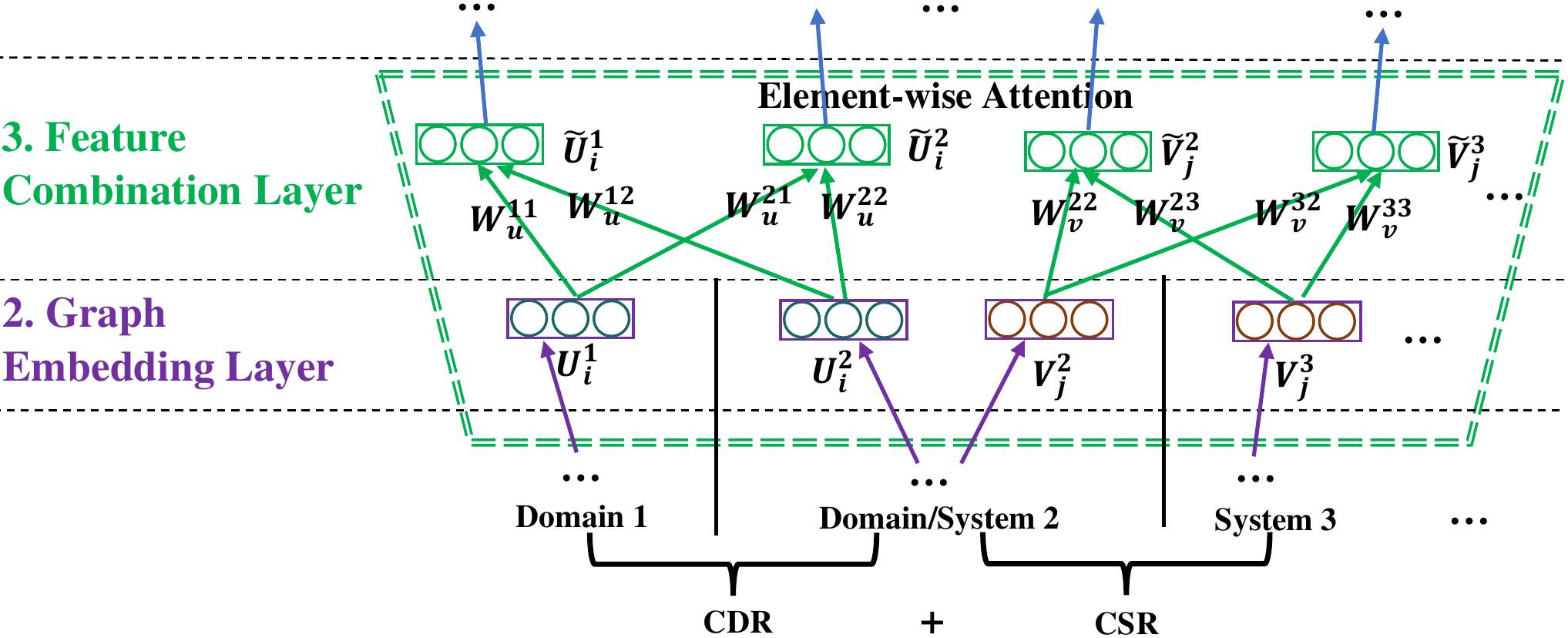}
		\caption{The overview of GA-CDR+CSR-P. Compared with GA-DTCDR-P, this is the extended structure for CDR+CSR scenario. Similar to GA-MTCDR-P, we ignore the same components with GA-DTCDR-P. Note that, for dataset (domain/system) $x$, $x \in\{1,2,...,a\}$, $\sum_{y=1}^{a} W^{xy}_u= 1$ and $\sum_{y=1}^{a} W^{xy}_v= 1$, where $a$ is the total number of datasets, $W^{xy}_u$ is the weight vector for the embedding of common users for CDR, and $W^{xy}_v$ is the weight vector for the embedding of common items for CSR. Domain/System $2$ contains common users with domain $1$ and common items with system $3$. The details are explained in Section \ref{subsubsection_GA-CDR+CSR}}
		\label{GeneralFramework_CDR+CSR}
	\end{center}
\end{figure}
\subsubsection{For GA-CDR+CSR-P} \label{subsubsection_GA-CDR+CSR} 
In GA-CDR+CSR-P (see Fig. \ref{GeneralFramework_CDR+CSR}), the element-wise attention mechanism is used to combine both the embeddings of common users for related domains and the embeddings of common items for related systems. Unlike GA-DTCDR-P and GA-MTCDR-P, in GA-CDR+CSR-P, if two or multiple datasets have common users, they should make cross-domain recommendations, thus these datasets are related domains to each other. While if two or multiple datasets have common items, they should make cross-system recommendations, thus these datasets are related systems to each other. For example, in Fig. \ref{GeneralFramework_CDR+CSR}, domain/system $2$ has the common users with domain $1$, thus it is a related domain for domain $1$. Meanwhile, domain/system $2$ has the common items with system $3$, thus it is also a related system for system $3$. Domain/system $2$ plays two roles, i.e., a related domain and a related system, in GA-CDR+CSR-P. 

In GA-CDR+CSR-P, there are $a$ related datesets $\mathcal{DS}$ (domains and systems). Similar to Eq. (\ref{Element-WiseAttention}), for a common user $u_i$ from dual/multiple domains $\mathcal{D}$ ($\mathcal{D} \in \mathcal{DS}$), his/her combined embedding $\tilde{U}^x_i$ for a domain $D^x$ ($D^x \in \mathcal{D}$) can be represented as:
\begin{equation}\label{Element-WiseAttention_CDR}
	\begin{aligned}
		&\tilde{U}^x_i=\sum_{D^y \in \mathcal{D}} W^{xy}_u \odot U^y_i, ~~~\sum_{D^y \in \mathcal{D}} W^{xy}_u =1.
	\end{aligned}
\end{equation}

Similarly, for a common item $v_j$ from dual/multiple systems $\mathcal{S}$ ($\mathcal{S} \in \mathcal{DS}$), its combined embedding $\tilde{V}^x_j$ for a system $S^x$ ($S^x \in \mathcal{S}$) can be represented as: 
\begin{equation}\label{Element-WiseAttention_CSR}
	\begin{aligned}
		&\tilde{V}^x_j=\sum_{S^y \in \mathcal{S}} W^{xy}_v \odot V^y_j, ~~~\sum_{S^y \in \mathcal{S}} W^{xy}_v =1,
	\end{aligned}
\end{equation}
where $W^{xy}_v$ is the weight vector of the embedding of common items from system $y$ for system $x$.

\subsection{Training for NN and Output Layers}
In this section, we introduce two training strategies, i.e., preliminary training and personalized training, for the neural network layers and output layer of our GA models. The preliminary training strategy is adopted by our preliminary work \cite{zhu2020graphical} and the personalized training strategy is adopted in this work (marked with `-P', e.g., GA-DTCDR-P).
\subsubsection{Preliminary Training}\label{Preliminary_training}
In our preliminary work \cite{zhu2020graphical}, we train our models with the following objective function in domain $x$:
\begin{equation}\label{ObjectiveFunction}
\begin{aligned}
&\min \limits_{P^x,Q^x,\Theta^x}{\sum \limits_{y \in Y^{x+} \cup Y^{x-}} { \ell(y , \hat{y})}+\lambda(\| P^x \|_{F}^2+\| Q^x \|_{F}^2)},\\
\end{aligned}
\end{equation}
where $\ell(y,\hat{y})$ is a loss function between an observed interaction $y$ and its corresponding predicted interaction $\hat{y}$ (see Eq. (\ref{LossFunction})), $Y^{x+}$ and $Y^{x-}$ denote all the observed and the unobserved user-item interactions in domain $x$ respectively, $\| P^x \|_{F}^2+\| Q^x\|_{F}^2$ is the regularizer (see Eq. (\ref{Embedding_Output})), $\lambda$ is a hyper-parameter which controls the importance of the regularizer, and $\Theta^x$ is the parameter set. To avoid our model over-fitted to $Y^+$ (positive instances), we randomly select a certain number of unobserved user-item interactions as negative instances, denoted by $Y^-_{sampled}$, to replace $Y^-$. This training strategy has been widely used in the existing approaches \cite{he2017neural}.

Unlike the unified loss functions in \cite{xin2015cross,jiang2016little,liu2020cross}, we train our recommendation model in each domain respectively and parallelly, which focuses on specifically improving the recommendation accuracy in each of the domains.

Based on rating information, the user-item interaction $y_{ij}$ between a user $u_i$ and an item $v_i$ can be represented as:
\begin{equation}\label{UI_Matrix}
y_{ij}=\left\{
\begin{aligned}
r_{ij}, ~~~~~~& {\rm if~}y_{ij} \in Y^+; \\
0,~~~~~~~ & {\rm if~} y_{ij} \in Y^-_{sampled};\\
null,~~~~~~~ & {\rm otherwise}.\\
\end{aligned}
\right.
\end{equation}

We choose a normalized cross-entropy loss which can be represented as:
\begin{equation}\label{LossFunction}
\ell(y , \hat{y}) =\frac{y}{max(R)} \log \hat{y} + (1-\frac{y}{max(R)}) \log (1-\hat{y}),
\end{equation}
where $max(R)$ is the maximum rating in a domain.

As shown in \emph{Neural Network Layers} of Fig. \ref{GeneralFramework_DTCDR}, our GA sub-frameworks employ a neural network, i.e., MLP, to represent a non-linear relationship between users and items. The input embedding matrices of users and items in domain $x$ for the MLP are $P^x_{in}=[\tilde{U}^x;U^{xd}]$ and $Q^x_{in}=V^x$ respectively, where $\tilde{U}^x$ is the combined embedding matrix of common users for domain $x$, and $U^{xd}$ is the embedding matrix of distinct users in domain $x$. Therefore the embedding of user $u_i$ and item embedding of item $v_j$ in the output layer of the MLP can be represented as:
\begin{equation}\label{Embedding_Output}
\begin{aligned}
&P_i^x=P_{{out}_i}^x=f(...f(f(P_{in_i}^x \cdot W_{P_1}^x) \cdot W_{P_2}^x)),\\
&Q_j^x=Q_{{out}_j}^x=f(...f(f(Q_{in_j}^x \cdot W_{Q_1}^x) \cdot W_{Q_2}^x)),
\end{aligned}
\end{equation}
where the activation function $f(*)$ is \emph{ReLU}, $W_{P_1}^x, W_{P_2}^x ...$ and $W_{Q_1}^x, W_{Q_2}^x ...$ are the weights of multi-layer networks in different layers in domain $x$ for $P^x_{in_i}$ and $Q^x_{in_j}$, respectively.

Finally, in \emph{Output Layer} of Fig. \ref{GeneralFramework_DTCDR}, the predicted interaction $\hat{y}_{ij}$ between $u_i$ and $v_j$ in domain $x$ is as follows:
\begin{equation}\label{PredictedInteraction}
\begin{aligned}
&\hat{y}_{ij}^x=cosine(P_i^x,Q_j^x)=\frac{P_i^x \cdot Q_j^x}{ \| P_i^x \| \| Q_j^x \|}.
\end{aligned}
\end{equation}
Compared with the conventional inner product, the biggest advantage of cosine distance for interaction prediction is that it does not need to normalize separately.

Similarly, we can train our models in each system.
\subsubsection{Personalized Training}\label{Personalized_training}
Although we have adopted the element-wise attention to combine the embeddings of common entities (users/items) from different datasets (domains/systems), our preliminary training strategy still suffers from the negative transfer problem. Especially in multi-target CDR and CDR+CSR scenarios, the recommendation performance may decline as more sparser datasets join in.

Inspired by the optimization problem in \cite{yang2020federated}, we propose a personalized objective function for our GA framework. This personalized training strategy first gives trainable weights on the pair-wise embedding differences of common entities between every two datasets, and then minimizes both local loss (see Eq. (\ref{LossFunction})) and these pair-wise embedding differences. The embeddings of common entities in different datasets tend to be similar but remain good personalization. Therefore, this personalized strategy can avoid negative transfer to some extent. The objective function in dataset $x$ is represented as follows:
\begin{equation}\label{ObjectiveFunction_PFL}
	\begin{aligned}
	&\min \limits_{P,Q,\Theta^x}{\sum \limits_{y \in Y^{x+} \cup Y^{x-}} {\ell(y , \hat{y})}+ \sum_{i \neq j}^{a}  \lambda^{ij}  A(\|W^{i} - W^{j}\|^2_F)},\\	
	\end{aligned}
\end{equation}
where $\ell(y , \hat{y})$ is the normalized corss-entropy loss (see Eq. (\ref{LossFunction})), $A$ is the attention-inducing function, which measures the embedding difference in a non-linear manner, $W^{i}$ ($P^{ic}$ or $Q^{ic}$) is the embeddings of common entities in dataset $i$. We adopt the negative exponential funciton, i.e., $1 - e^{-\|W^{i} - W^{j}\|^2_F/\theta}$ with a hyper-parameter $\theta$, which has been widely-used in many existing personalized approaches \cite{huang2020personalized}. Additionally, the existing personalized approaches tend to choose a fixed hyper-parameter $\lambda$ to control the weight on embedding difference. But in our objective function, we use a set of trainable variables ($ \sum_{i \neq j}^{a}  \lambda^{ij} =1$, e.g., $\lambda^{ij}$ is the weight for the embedding difference between datasets $i$ and $j$), to train suitable weights on the embedding differences. These trainable weights can effectively control the importance of pair-wise embedding difference of common entities and thus the trained recommendation model in each dataset can achieve good personalization. Therefore, our GA-DTCDR-P, GA-MTCDR-P, and GA-CDR+CSR-P can alleviate negative transfer by using this personalized training strategy.

\section{Experiments and Analysis} \label{Experiments&Analysis}
We conduct extensive experiments on four real-world datasets to answer the following key questions: 
\begin{itemize}[leftmargin=*] \setlength{\itemsep}{-\itemsep}
\item \textbf{Q1:} How do our GA models (GA-DTCDR-P, GA-MTCDR-P, and GA-CDR+CSR-P) perform when compared with the state-of-the-art models (see Result 1)? 
\item \textbf{Q2:} How do the element-wise attention mechanism and personalized training strategy contribute to performance improvement (see Result 2)? 
\item \textbf{Q3:} How does the dimension $k$ of embeddings affect the performance of our models (see Result 3)? 
\item \textbf{Q4:} How do our models perform on Top-$N$ recommended lists (see Result 4)?
\item \textbf{Q5:} How do the data sparsity and the scale of overlap affect the performance of our models (see Result 5)?
\end{itemize}

\subsection{Experimental Settings}
\begin{table}[t]
	\begin{center}
		\caption{Experimental datasets and tasks} \label{Datasets}
		\begin{tabular}{c|c|c|c|c}
			\hline
			\textbf{Datasets}& \multicolumn{3}{c|}{Douban}& MovieLens\\
			\hline
			\textbf{Domains}   & Book & Music & Movie  & Movie\\
			\hline
			\textbf{\#Users} & 2,110 & 1,672 & 2,712 & 10,000\\
			\hline
			\textbf{\#Items} & 6,777 & 5,567    & 34,893 & 9,395\\
			\hline
			\textbf{\#Interactions} & 96,041 & 69,709  & 1,278,401 & 1,462,905\\
			\hline
			\textbf{Density}& 0.67\% & 0.75\%  & 1.35\% & 1.56\%\\
			\hline
			\hline
		\end{tabular}
	\scriptsize{
		\begin{tabular}{c|c|c@{~}|@{~}c@{~}|@{~}c@{~}}
				\hline
				\multicolumn{2}{c@{~}|}{\textbf{Tasks}}&\textbf{Sparser}&\textbf{Richer}& \textbf{Overlap}\\
				\hline
				\multirow{2}{*}{CDR}&\textbf{Task 1} & DoubanBook& DoubanMovie & \#Common Users = 2,106\\
				\cline{2-5}
				&\textbf{Task 2} & DoubanMusic & DoubanMovie & \#Common Users = 1,666\\
				\hline
				CSR & \textbf{Task 3}& DoubanMovie & MovieLens  & \#Common Items = 4,115\\
				\hline
				\hline
			\end{tabular}
			\begin{tabular}{c|@{~}c@{~}|@{~}c@{~}}
				\hline
				\multicolumn{2}{c@{~}|@{~}}{\textbf{Tasks}}&\textbf{Domains/Systems}\\
				\hline
				MTCDR&\textbf{Task 4} & \makecell{DoubanBook+DoubanMusic+DoubanMovie \\ \#Common Users = 1,662}\\
				\hline
				CDR+CSR&\textbf{Task 5} &  \makecell{DoubanBook+DoubanMovie+MovieLens \\ \#Common Users (DoubanBook+DoubanMovie)= 2,106 \\ \#Common Items (DoubanMovie+MovieLens) = 4,115}\\
				\hline
			\end{tabular}
		}
	\end{center}
\end{table}
\begin{table*}[t]
	\begin{center}
		\caption{The comparison of the baselines and our methods} \label{Baselines}
		\resizebox{\textwidth}{!}{
			\begin{tabular}{c@{   }|@{   }c@{   }|@{   }c@{   }|@{   }c@{   }|@{   }c@{   }|@{   }c@{   }|@{   }c@{   }}
				\hline
				\multicolumn{3}{c@{   }|@{   }}{\textbf{Model}}& \textbf{Training Data}& \makecell{\textbf{Encoding}} & \makecell{\textbf{Embedding}}& \makecell{\textbf{Transfer Strategy}} \\
				\hline
				\hline
				\multirow{10}{*}{Baselines}&\multirow{2}{*}{\makecell{Single-Domain\\ Recommendation (SDR)}}&\textbf{NeuMF} \cite{he2017neural}& Rating & One-hot & Non-linear MLP& -\\ \cline{3-7}
				&&\textbf{DMF} \cite{xue2017deep}& Rating& Rating Vector &Non-linear MLP & -\\ \cline{2-7}
				&\multirow{4}{*}{\makecell{Single-Target \\Cross-Domain\\ Recommendation \\ (CDR)}}& \textbf{CTR-RBF} \cite{xin2015cross}& Rating \& Content & Topic Modeling & Linear MF & \makecell{Mapping \& \\Transfer Learning}\\ \cline{3-7}
				&& \makecell{\textbf{BPR\_DCDCSR} \cite{zhu2018deep}}& Rating & Random Initialization & Linear MF & Combination \& MLP\\ \cline{3-7}
				&& \makecell{\textbf{TMH} \cite{hu2019transfer}}& Rating \& Content & One-hot & Non-linear MLP & \makecell{Mapping \& Transfer\\ Learning \& Attention} \\ \cline{2-7}
				& \multirow{3}{*}{Dual-Target CDR} & \makecell{\textbf{DMF\_DTCDR\_Concat} \cite{zhu2019dtcdr}}& Rating \& Content & Rating Vector & Non-linear MLP & \makecell{Multi-task Learning \\ \& Concatenation} \\
				\cline{3-7}
				&&\makecell{\textbf{DDTCDR} \cite{li2019ddtcdr}}& Rating & One-hot \& Multi-hot & Non-linear MLP & \makecell{Dual Transfer Learning} \\ 
				\hline
				\hline
				\multirow{12}{*}{\makecell{Our\\ Methods}}&\multirow{6}{*}{\makecell{Dual-Target CDR}}&\makecell{\textbf{GA-DTCDR\_Average} \cite{zhu2020graphical}\\ (a variant of GA-DTCDR \\for ablation study)}& Rating \& Content & Heterogeneous Graph & Graph Embedding & \makecell{Combination \\ (Average-Pooling)}\\ \cline{3-7}
				&& \makecell{\textbf{GA-DTCDR} \cite{zhu2020graphical}\\ (our prior work, \\preliminary training)} & Rating \& Content & Heterogeneous Graph & Graph Embedding &\makecell{Combination \\(Element-wise Attention)}\\ 
				\cline{3-7}
				&& \makecell{\textbf{GA-DTCDR-P} \\ (personalized training)} & Rating \& Content & Heterogeneous Graph & Graph Embedding & \makecell{\textbf{Element-wise Attention} \\ \& \textbf{Personalization}}\\
				\cline{2-7} 
				&Multi-Target CDR&\makecell{\textbf{GA-MTCDR-P} \\ (personalized training)}& Rating \& Content & Heterogeneous Graph & Graph Embedding &\makecell{\textbf{Element-wise Attention} \\ \& \textbf{Personalization}}\\ 
				\cline{2-7}
				&CDR+CSR&\makecell{\textbf{GA-CDR+CSR-P} \\(personalized training)}& Rating \& Content & Heterogeneous Graph & Graph Embedding &\makecell{\textbf{Element-wise Attention} \\ \& \textbf{Personalization}}\\ 
				\hline
			\end{tabular}
		}
	\end{center}
\end{table*}
\subsubsection{Experimental Datasets and Tasks} 
To validate the recommendation performance of our GA approaches and baseline approaches, we choose four real-world datasets, i.e., three Douban subsets (DoubanBook, DoubanMusic, and DoubanMovie) \cite{zhu2019dtcdr}, and MovieLens 20M \cite{harper2016movielens}. For the three Douban subsets, we retain the users and items with at least $5$ interactions each user, while for MovieLens 20M, we extract a MovieLens subset containing $10,000$ users with at least $5$ interactions each user as well. This filtering strategy has been widely used in the existing approaches \cite{yuan2019darec,zhu2019dtcdr}. The three Douban subsets contain ratings, reviews, tags, user profiles, and item details while MovieLens contains ratings, tags, and item details. Based on these four datasets, we design two CDR tasks (\textbf{Task 1} \& \textbf{2} in Table \ref{Datasets}) and one CSR task (\textbf{Task 3}) to validate the recommendation performance in dual-target CDR and CSR scenarios, respectively. In addition, we design one MTCDR task (\textbf{Task 4}) and one CDR+CSR task (\textbf{Task 5}) to validate the recommendation performance in multi-target CDR and CDR+CSR scenarios, respectively. We list the dataset statistics and designed tasks in Table \ref{Datasets}. 

\begin{table*}[t]
	\begin{center}
		\caption{The experimental results (HR@$10$ \& NDCG@$10$) for Tasks 1, 2, and 3 (the best-performing baselines with results marked with \textbf{*} while our best-performing models with results marked with black body)} \label{ExperimentalResults}
		\resizebox{\textwidth}{!}{
			\begin{tabular}{@{}c@{}|@{}c@{}||@{}c@{}|@{}c@{}||@{}c@{}|@{}c@{}|@{}c@{}||@{}c@{}|@{}c@{}||@{}c@{}|@{}c@{}||@{}c@{}||@{}c@{}}
				\hline
				\multirow{3}{*}{\textbf{Task}} & \multirow{3}{*}{\makecell{\textbf{Domain} \\ (R: Richer \\ S: Sparser)}}& \multicolumn{2}{c@{}||@{}}{\textbf{\makecell{SDR Baselines}}} & \multicolumn{3}{c@{}||@{}}{\makecell{\textbf{Single-Target CDR Baselines}}}&\multicolumn{2}{c@{}||@{}}{\makecell{\textbf{DTCDR Baselines}}} & \multicolumn{2}{c@{}||@{}}{\makecell{\textbf{Our DTCDR}\\\textbf{(our prior works)}}}& \makecell{\textbf{Our DTCDR}}&  \multirow{2}{*}{\makecell{\textbf{Improvement} \\ \textbf{(GA-DTCDR-P} \\vs. \textbf{best baselines)}}} \\
				\cline{3-12}
				&& NeuMF & DMF & CTR-RBF & \makecell{BPR\\\_DCDCSR} & TMH & \makecell{DMF\_DTCDR\\\_Concat} & DDTCDR& \makecell{GA-DTCDR\\\_Average}& GA-DTCDR& \textbf{\makecell{GA-DTCDR-P}}& \\
				\cline{3-12}
				&&HR~NDCG& HR~NDCG &HR~NDCG& HR~NDCG&HR~NDCG & HR~NDCG & HR~NDCG & HR~NDCG & HR~NDCG& HR~NDCG & HR~NDCG\\
				\hline
				\hline
				\multirow{2}{*}{\makecell{Task1\\ ($k=8$)}}&DoubanBook (S) &.3810~.2151&.3841~.2265&.3830~.2217&.3954~.2419&.4199~\textbf{.2583*}&\textbf{.4412*}~.2571&.4033~.2257&.4057~.2513&.4479~.2759&\textbf{.4481}~\textbf{.2766}&1.56\% ~7.08\% \\
				\cline{2-13}
				&DoubanMovie (R)&.5266~.2911&.5498~.3114&-~~-&-~~-&-~~-&\textbf{.6032*}~\textbf{.3732*}&.5612~.3185&.5968~.3546&.6518~.4025&\textbf{.6536}~\textbf{.4059}&8.35\% ~8.76\% \\
				\hline
				\multirow{2}{*}{\makecell{Task 1\\ ($k=16$)}}&DoubanBook (S) &.3833~.2181&.3854~.2356&.3870~.2256&.4014~.2413&.4331~\textbf{.2522*}&\textbf{.4408*}~.2513&.4054~.2292&.4190~.2577&.4706~.2900&\textbf{.4718}~\textbf{.2909}&7.03\% ~15.34\% \\
				\cline{2-13}
				&DoubanMovie (R)&.5282~.2939&.5573~.3141&-~~-&-~~-&-~~-&\textbf{.6080*}~\textbf{.3721*}&.5750~.3595&.6013~.3596&.6566~.4014&\textbf{.6582}~\textbf{.4043}&8.26\% ~8.65\% \\
				\hline
				\multirow{2}{*}{\makecell{Task 1\\ ($k=32$)}}&DoubanBook (S) &.3899~.2182&.3871~.2340&.3956~.2264&.4079~.2436&\textbf{.4468*}~\textbf{.2647*}&.4318~.2461&.4180~.2344&.4346~.2610&.4758~.2896&\textbf{.4771}~\textbf{.2899}&6.78\% ~9.52\% \\
				\cline{2-13}
				&DoubanMovie (R)&.5411~.2991&.5612~.3254&-~~-&-~~-&-~~-&\textbf{.6011*}~\textbf{.3718*}&.5739~.3386&.6374~.3896&\textbf{.6747}~.4187&.6742~\textbf{.4277}&12.16\% ~15.03\% \\
				\hline
				\multirow{2}{*}{\makecell{Task 1\\ ($k=64$)}}&DoubanBook (S) &.3908~.2226&.3917~.2362&.4017~.2314&.4107~.2454&\textbf{.4504*~.2768*}&.4265~.2452&.4258~.2430&.4423~.2671&.4882~.3026&\textbf{.4891}~\textbf{.3131}&8.60\% ~13.11\% \\
				\cline{2-13}
				&DoubanMovie (R)&.5449~.3152&.5632~.3387&-~~-&-~~-&-~~-&\textbf{.5998*~.3649*}&.5825~.3553&.6416~.3941&\textbf{.6817}~.4205&.6802~\textbf{.4249}&13.40\% ~16.44\% \\
				\hline
				\multirow{2}{*}{\makecell{Task 1\\ ($k=128$)}}&DoubanBook (S) &.4012~.2310&.4046~.2451&.4171~.2532&.4111~.2431&\textbf{.4523*~.2814*}&.4317~.2510&.4225~.2439&.4490~.2691&.4995~.3098&\textbf{.5011}~\textbf{.3121}&10.79\% ~10.91\% \\
				\cline{2-13}
				&DoubanMovie (R)&.5512~.3301&.5776~.3505&-~~-&-~~-&-~~-&\textbf{.5991*~.3680*}&.5863~.3589&.6449~.3981&\textbf{.6957}~\textbf{.4406}&.6942~.4391&15.87\% ~19.32\% \\
				\hline
				\hline
				\hline
				\multirow{2}{*}{\makecell{Task 2\\ ($k=8$)}}&DoubanMusic (S) &.3135~.1703&.3127~.1812&.3227~.1895&.3259~.1894&.3579~.2034&\textbf{.3614*~.2117*}&.3302~.1930&.3690~.2109&.3852~.2166&\textbf{.3871}~\textbf{.2231}&7.11\% ~5.38\% \\
				\cline{2-13}
				&DoubanMovie (R)&.5266~.2911&.5498~.3114&-~~-&-~~-&-~~-&\textbf{.5873*~.3867*}&.5655~.3629&.5987~.3731&.6470~.3983&\textbf{.6473}~\textbf{.4008}&10.22\% ~3.65\% \\
				\hline
				\multirow{2}{*}{\makecell{Task 2\\ ($k=16$)}}&DoubanMusic (S) &.3190~.1731&.3170~.1891&.3121~.1761&.3261~.1901&.3612~.2137&\textbf{.3663*~.2213*}&.3451~.2092&.3706~.2037&.3947~.2256&\textbf{.3976}~\textbf{.2330}&8.54\% ~5.29\% \\
				\cline{2-13}
				&DoubanMovie (R)&.5282~.2939&.5573~.3141&-~~-&-~~-&-~~-&\textbf{.5887*~.3863*}&.5704~.3676&.6058~.3716&.6426~.3950&\textbf{.6463}~\textbf{.4001}&9.78\% ~3.57\% \\
				\hline
				\multirow{2}{*}{\makecell{Task 2\\ ($k=32$)}}&DoubanMusic (S) &.3198~.1771&.3218~.1912&.3141~.1844&.3271~.1931&\textbf{.3701*~.2202*}&.3607~.2201&.3463~.2050&.3789~.2056&.4133~.2318&\textbf{.4165}~\textbf{.2449}&12.53\% ~11.22\% \\
				\cline{2-13}
				&DoubanMovie (R)&.5411~.2991&.5612~.3254&-~~-&-~~-&-~~-&\textbf{.5770*~.3758*}&.5739~.3726&.6145~.3754&\textbf{.6677}~\textbf{.4141}&.6672~.4121&15.63\% ~9.66\% \\
				\hline
				\multirow{2}{*}{\makecell{Task 2\\ ($k=64$)}}&DoubanMusic (S) &.3242~.1791&.3267~.1926&.3324~.1916&.3304~.2001&\textbf{.3882*~.2323*}&.3571~.2109&.3466~.2045&.3812~.2144&.4384~.2489&\textbf{.4402}~\textbf{.2527}&13.40\% ~8.78\% \\
				\cline{2-13}
				&DoubanMovie (R)&.5449~.3152&.5632~.3387&-~~-&-~~-&-~~-&\textbf{.5787*~.3705*}&.5719~.3621&.6120~.3681&\textbf{.6817}~.4284&.6811~\textbf{.4289}&17.69\% ~15.76\% \\
				\hline
				\multirow{2}{*}{\makecell{Task 2\\ ($k=128$)}}&DoubanMusic (S) &.3314~.1810&.3301~.1971&.3412~.1954&.3452~.2074&\textbf{.3946*~.2430*}&.3580~.2132&.3520~.2117&.3996~.2207&.4491~.2604&\textbf{.4496}~\textbf{.2669}&13.94\% ~9.84\% \\
				\cline{2-13}
				&DoubanMovie (R)&.5512~.3301&.5776~.3505&-~~-&-~~-&-~~-&\textbf{.5792*}~.3742&.5748~\textbf{.3762*}&.6311~.3859&\textbf{.7068}~.4526&.7053~\textbf{.4533}&21.77\% ~20.49\% \\
				\hline
				\hline
				\hline
				\multirow{2}{*}{\makecell{Task 3\\ ($k=8$)}}&DoubanMovie (S) &.5266~.2911&.5498~.3114&.5514~.3156&.5762~.3347&.5987~.3487&\textbf{.6387*~.3628*}&.6070~.3522&.6140~.3572&.6486~.4005&\textbf{.6491}~\textbf{.4032}&16.28\% ~11.34\% \\
				\cline{2-13}
				&MovieLens (R)&.7818~.5024&.8115~.5219&-~~-&-~~-&-~~-&\textbf{.8328*~.5293*}&.8211~.5283&.8225~.5241&.8541~.5372&\textbf{.8584}~\textbf{.5388}&3.07\% ~1.79\% \\
				\hline
				\multirow{2}{*}{\makecell{Task 3\\ ($k=16$)}}&DoubanMovie (S) &.5282~.2939&.5573~.3141&.5631~.3213&.5816~.3438&.6031~.3580&\textbf{.6391*~.3606*}&.6100~.3518&.6266~.3710&.6514~.4018&\textbf{.6526}~\textbf{.4056}&2.11\% ~12.48\% \\
				\cline{2-13}
				&MovieLens (R)&.7901~.5084&.8143~.5212&-~~-&-~~-&-~~-&\textbf{.8312*~.5260*}&.8263~.5170&.8280~.5277&\textbf{.8547}~\textbf{.5381}&.8542~.5376&2.76\% ~2.21\% \\
				\hline
				\multirow{2}{*}{\makecell{Task 3\\ ($k=32$)}}&DoubanMovie (S) &.5411~.2991&.5612~.3254&.5721~.3347&.5821~.3447&.6108~\textbf{.3733*}&\textbf{.6530*}~.3631&.6137~.3460&.6310~.3776&.6598~.4087&\textbf{.6603}~\textbf{.4123}&1.11\% ~10.45\% \\
				\cline{2-13}
				&MovieLens (R)&.7978~.5124&.8180~\textbf{.5231*}&-~~-&-~~-&-~~-&\textbf{.8243*}~.5213&.8111~.5167&.8301~.5280&.8612~.5478&\textbf{.8614}~\textbf{.5488}&4.50\% ~4.91\% \\
				\hline
				\multirow{2}{*}{\makecell{Task 3\\ ($k=64$)}}&DoubanMovie (S) &.5449~.3152&.5632~.3387&.5704~.3327&.5926~.3559&.6186~\textbf{.3754*}&\textbf{.6477*}~.3605&.6200~.3544&.6423~.3841&.6654~.4101&\textbf{.6665}~\textbf{.4158}&2.90\% ~10.76\% \\
				\cline{2-13}
				&MovieLens (R)&.7935~.5149&\textbf{.8231*}~.5277&-~~-&-~~-&-~~-&.8200~\textbf{.5382*}&.8130~.5198&.8324~.5320&\textbf{.8668}~.5516&.8654~\textbf{.5532}&5.14\% ~2.79\% \\
				\hline
				\multirow{2}{*}{\makecell{Task 3\\ ($k=128$)}}&DoubanMovie (S) &.5512~.3301&.5776~.3505&.5912~.3741&.6142~.3904&.6314~\textbf{.3927*}&\textbf{.6521*}~.3642&.6222~.3714&.6489~.3792&.6812~.4198&\textbf{.6838}~\textbf{.4312}&4.86\% ~9.80\% \\
				\cline{2-13}
				&MovieLens (R)&.8042~.5205&\textbf{.8319*}~.5344&-~~-&-~~-&-~~-&.8267~\textbf{.5401*}&.8210~.5311&.8349~.5381&.8642~.5512&\textbf{.8651}~\textbf{.5553}&3.99\% ~2.81\% \\
				\hline
			\end{tabular}
		}
	\end{center}
\end{table*}

\subsubsection{Parameter Setting} 
For a fair comparison, we optimize the parameters of our GA-DTCDR-P, GA-MTCDR-P, GA-CDR+CSR-P, and those of the baselines according to the parameter settings in their original papers. For \emph{Graph Embedding Layer} of GA framework, we set the hyper-parameters of Doc2vec and Node2vec models as suggested in \cite{le2014distributed,grover2016node2vec}, and the sampling probability $\alpha$ as $0.05$. In \emph{Neural Network Layers} of GA framework, the structure of the layers is `$k \rightarrow 2k \rightarrow 4k \rightarrow 8k \rightarrow 4k \rightarrow 2k \rightarrow k$', the parameters of the neural network are initialized as the Gaussian distribution $X \sim \mathcal{N}(0,0.01)$. For training our GA-DTCDR-P, GA-MTCDR-P, and GA-CDR+CSR-P, we randomly select $7$ negative instances for each observed positive instance into $Y^-_{sampled}$, adopt Adam \cite{kingma2014adam} to train the neural network, and set the maximum number of training epochs to $50$. The learning rate is $0.001$, the regularization coefficient $\lambda$ is $0.001$, and the batch size is $1,024$. To answer \textbf{Q3}, the dimension $k$ of the embedding varies in $\{8,16,32,64,128\}$.

\subsubsection{Evaluation Metrics} To evaluate the recommendation performance of our GA-DTCDR-P, GA-MTCDR-P, GA-CDR+CSR-P models, and baseline models, we adopt the ranking-based evaluation strategy, i.e., \emph{leave-one-out evaluation}, which has been widely used in the literature \cite{xue2017deep,wang2019kgat}. For each test user, we choose the latest interaction with a test item as the test interaction and randomly sample $99$ unobserved interactions for the test user, and then rank the test item among the $100$ items. \emph{Leave-one-out evaluation} includes two main metrics, i.e., \emph{Hit Ratio (HR)} and \emph{Normalized Discounted Cumulative Gain (NDCG)} \cite{wang2019kgat}. \emph{HR}@$N$ is the recall rate while \emph{NDCG}@$N$ measures the specific ranking quality that assigns high scores to hits at top position ranks. Note that we only report \emph{HR}@$10$ and \emph{NDCG}@$10$ results in \textbf{Results 1-3}, and \emph{HR}@$N$ and \emph{NDCG}@$N$ results in \textbf{Result 4}.

\subsubsection{Comparison Methods} As shown in Table \ref{Baselines}, we compare our GA models with seven baseline models in three groups, i.e., (1) Single-Domain Recommendation (SDR), (2) Single-Target Cross-Domain Recommendation (CDR), and (3) Dual-Target CDR. All seven baselines are representative and/or state-of-the-art approaches for each group. Also, for the ablation study, in addition to GA-DTCDR and GA-DTCDR-P, we implement a simplified version of GA-DTCDR, i.e., GA-DTCDR\_Average (replacing element-wise attention with a fixed combination strategy, i.e., average-pooling). For a clear comparison, in Table \ref{Baselines}, we list the detailed training data types, encoding strategies, embedding strategies, and transfer strategies of all the models implemented in the experiments.

\begin{table}[t]
	\caption{The experimental results of GA-MTCDR-P for Task 4}
	\label{ExperimentalResults_Task4}
	\begin{center}
		\resizebox{\columnwidth}{!}{
			\begin{tabular}{@{  }c@{  }|@{  }c@{  }|@{  }c@{  }|@{  }c@{  }|@{  }c@{  }|@{  }c@{  }}
				\hline
				\multirow{2}{*}{\textbf{Domain}}&\textbf{$k=8$}&\textbf{$k=16$}&\textbf{$k=32$}&\textbf{$k=64$}&\textbf{$k=128$}\\
				\cline{2-6}
				&HR~NDCG & HR~NDCG & HR~NDCG & HR~NDCG& HR~NDCG\\
				\hline
				DoubanBook&.4493~.2813&.4721~.2913&.4805~.3084&.4903~.3158&.5022~.3161\\
				\hline
				DoubanMusic&.3889~.2274&.3995~.2419&.4174~.2452&.4377~.2598&.4429~.2671\\
				\hline
				DoubanMoive&.6496~.4042&.6521~.4041&.6761~.4280&.6831~.4314&.7064~.4542\\
				\hline
			\end{tabular}

		}
	\end{center}
\end{table}
\begin{table}[t]
	\caption{The experimental results GA-CDR+CSR-P for Task 5} \label{ExperimentalResults_Task5}
	\begin{center}
		\resizebox{\columnwidth}{!}{
			\begin{tabular}{@{    }c@{    }|@{    }c@{    }|@{    }c@{    }|@{    }c@{    }|@{    }c@{    }|@{    }c@{    }}
				\hline
				\multirow{2}{*}{\textbf{Domain/System}}&\textbf{$k=8$}&\textbf{$k=16$}&\textbf{$k=32$}&\textbf{$k=64$}&\textbf{$k=128$}\\
				\cline{2-6}
				&HR~NDCG & HR~NDCG & HR~NDCG & HR~NDCG& HR~NDCG\\
				\hline
				DoubanBook&.4503~.2794&.4701~.2877&.4781~.3063&.4884~.3121&.4991~.3187\\
				\hline
				DoubanMoive&.6493~.4026&.6517~.4033&.6595~.4098&.6750~.4272&.6916~.4435\\
				\hline
				MoiveLens&.8512~.5249&.8533~.5303&.8594~.5486&.8685~.5536&.8632~.5527\\
				\hline
			\end{tabular}		
		}
	\end{center}
\end{table}
\subsection{Performance Comparison and Analysis}
\begin{figure*}[t]
	\centering
	\includegraphics[width=0.75\textwidth]{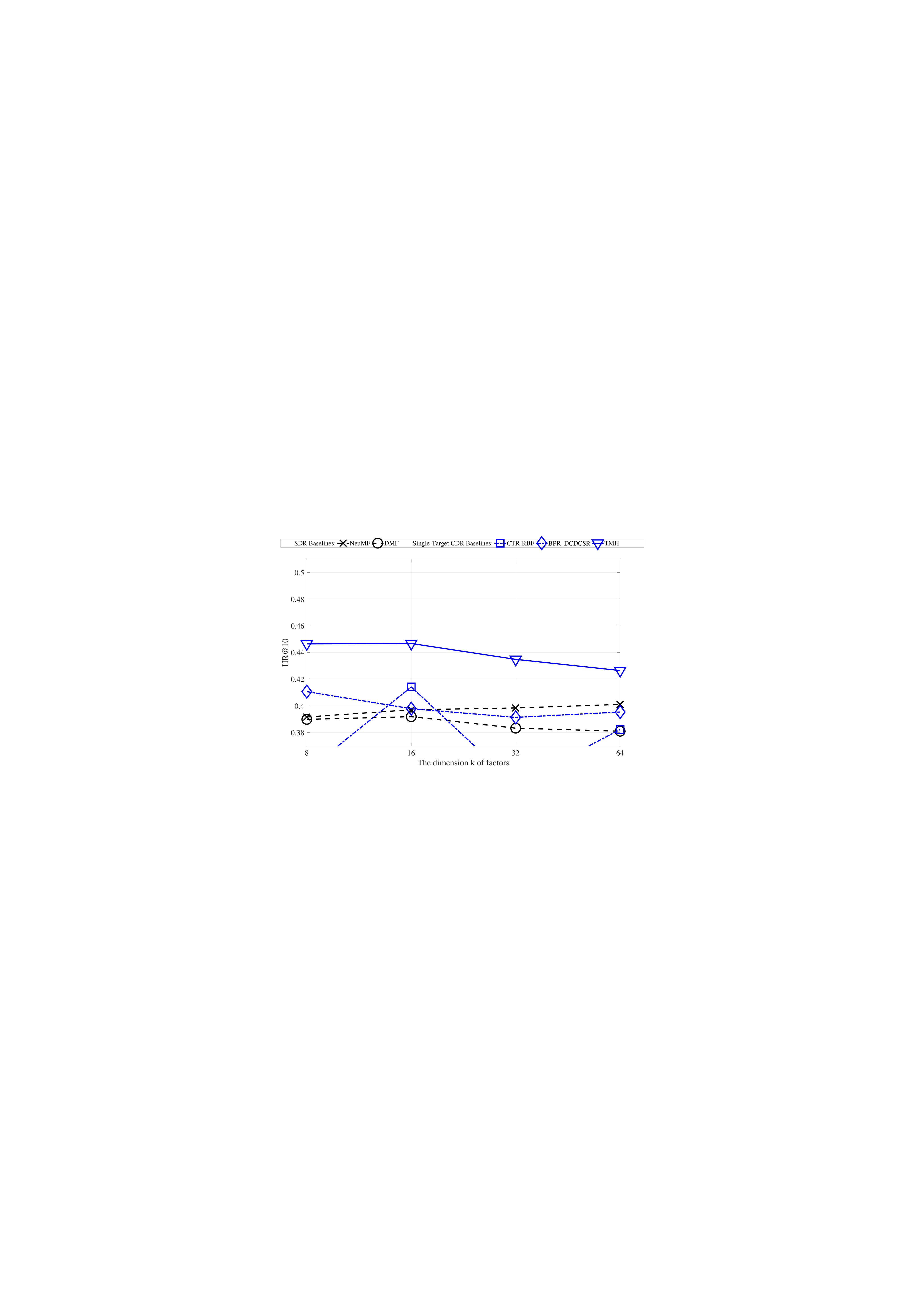}\\
	\includegraphics[width=0.9\textwidth]{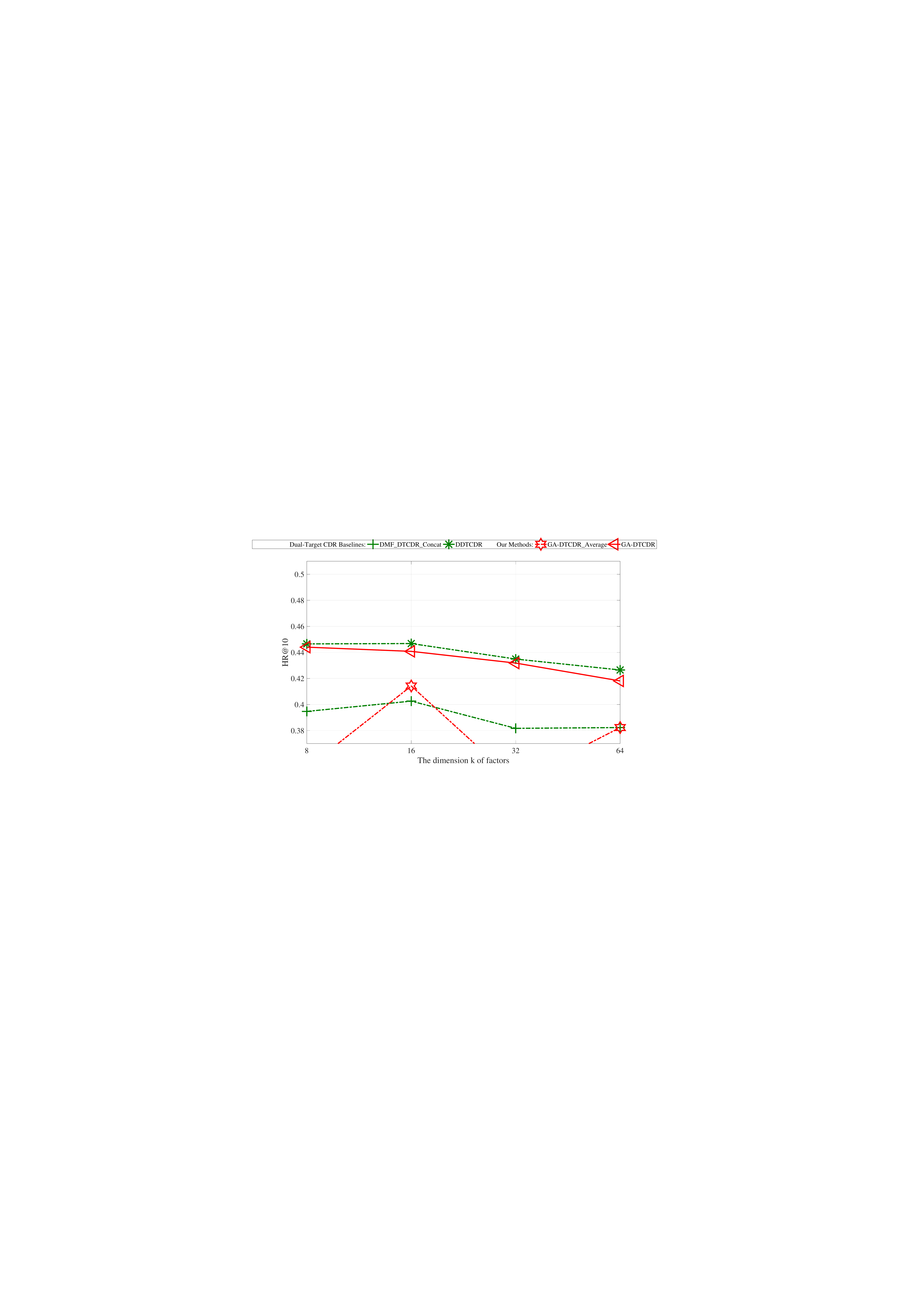}\\
	\subfigure[DoubanBook (HR@$N$)]{
		\includegraphics[width=0.24\textwidth]{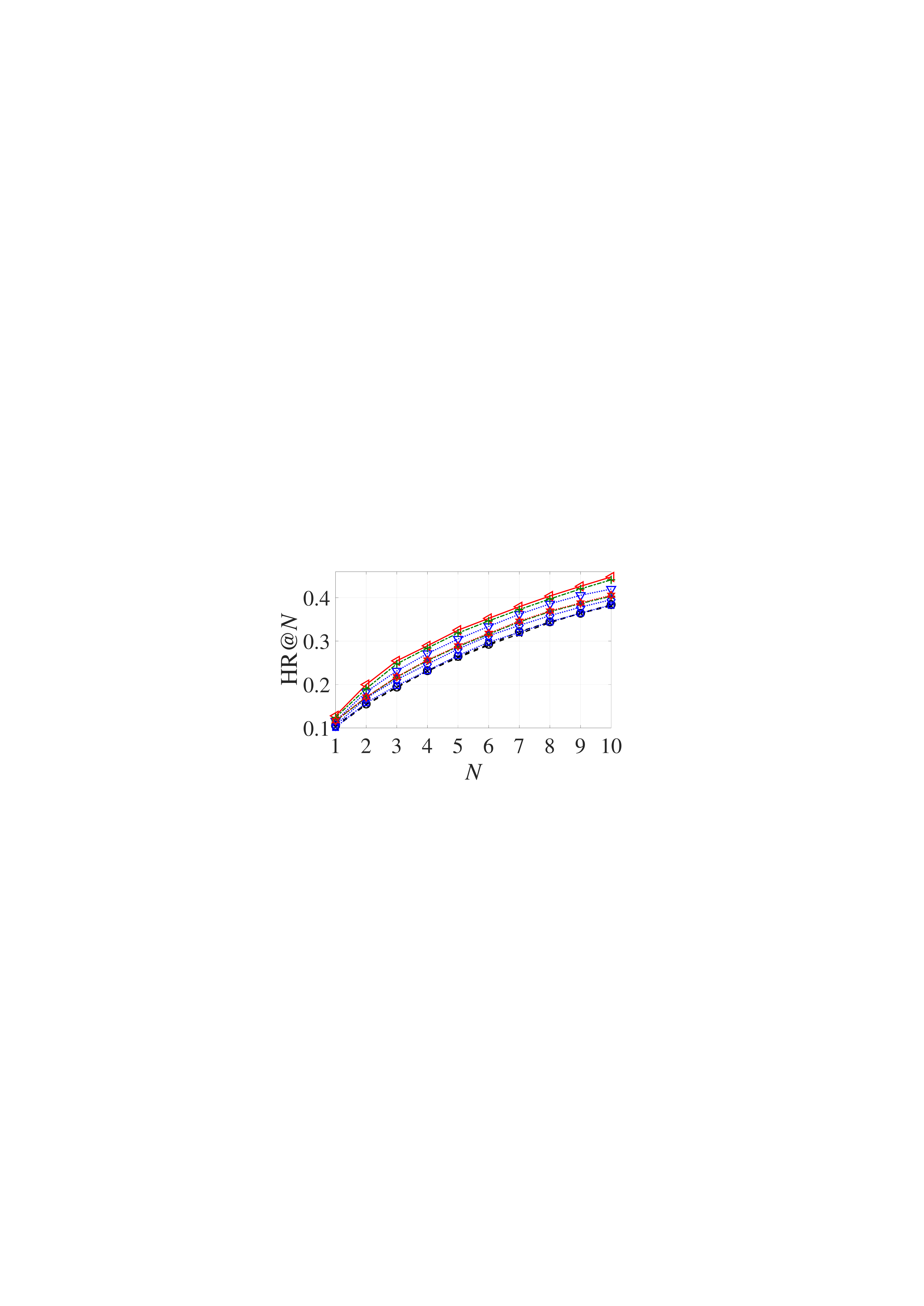}}
	\subfigure[DoubanBook (NDCG@$N$)]{
		\includegraphics[width=0.24\textwidth]{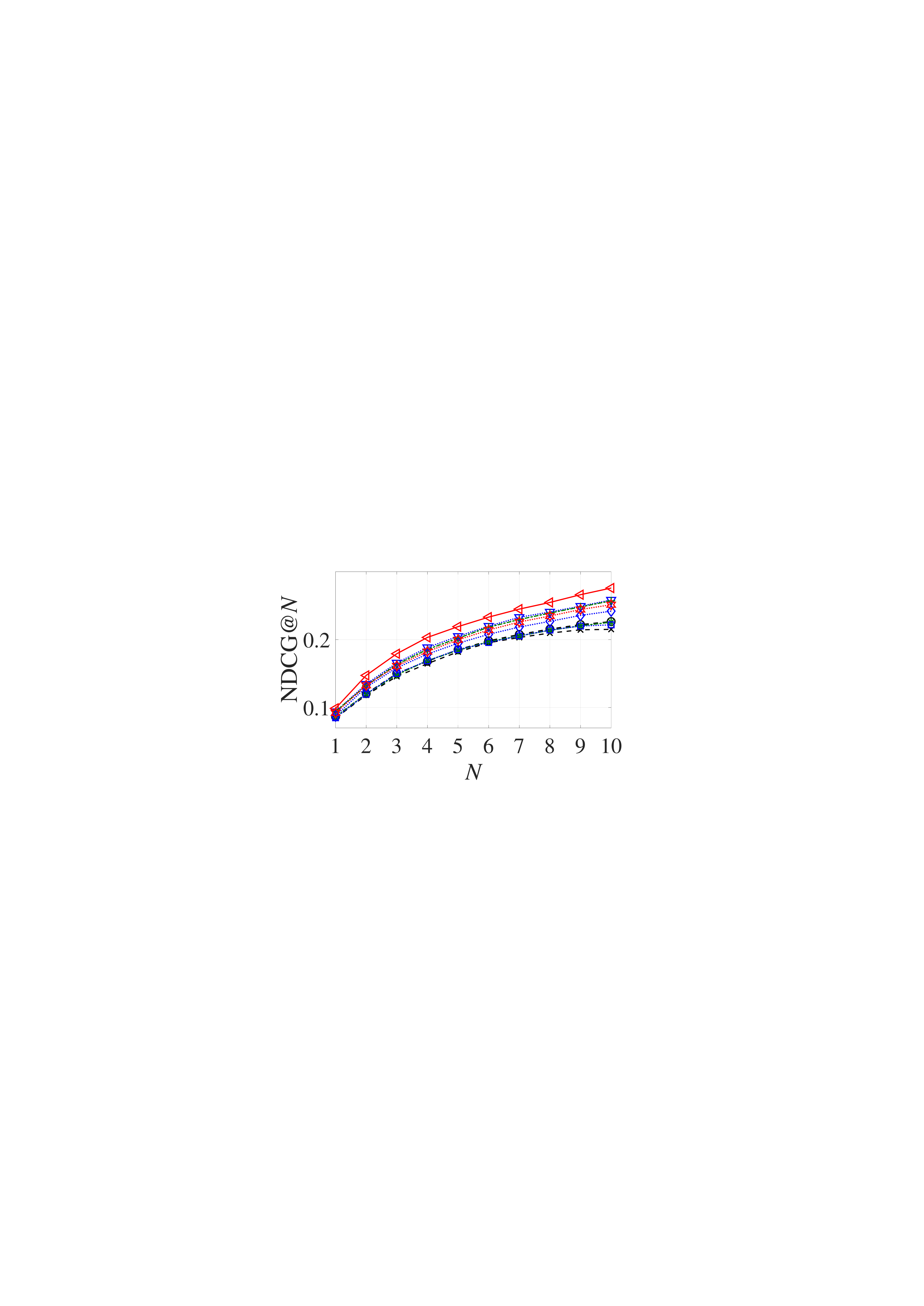}}
	\subfigure[DoubanMovie (HR@$N$)]{
		\includegraphics[width=0.24\textwidth]{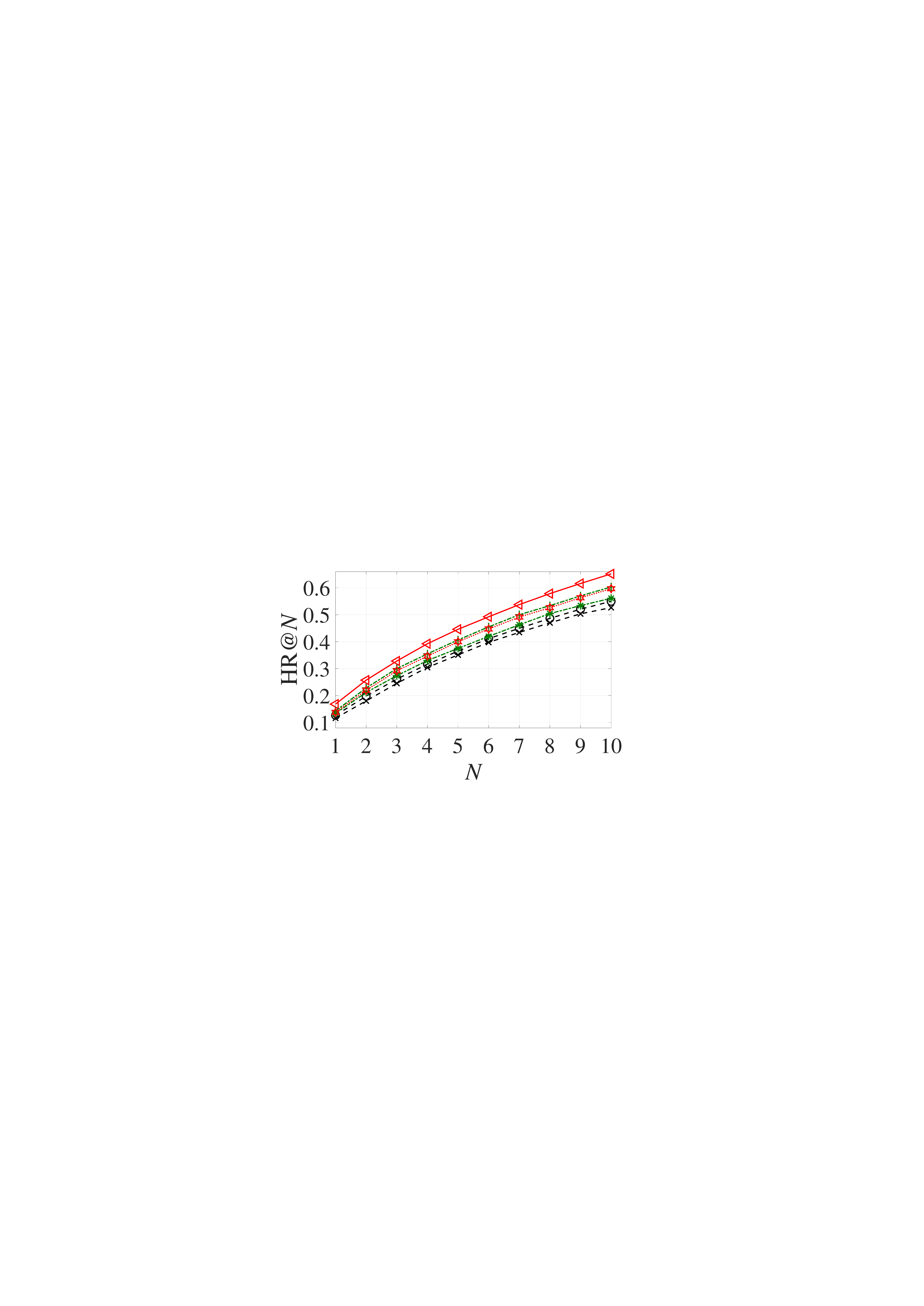}}
	\subfigure[DoubanMovie (NDCG@$N$)]{
		\includegraphics[width=0.24\textwidth]{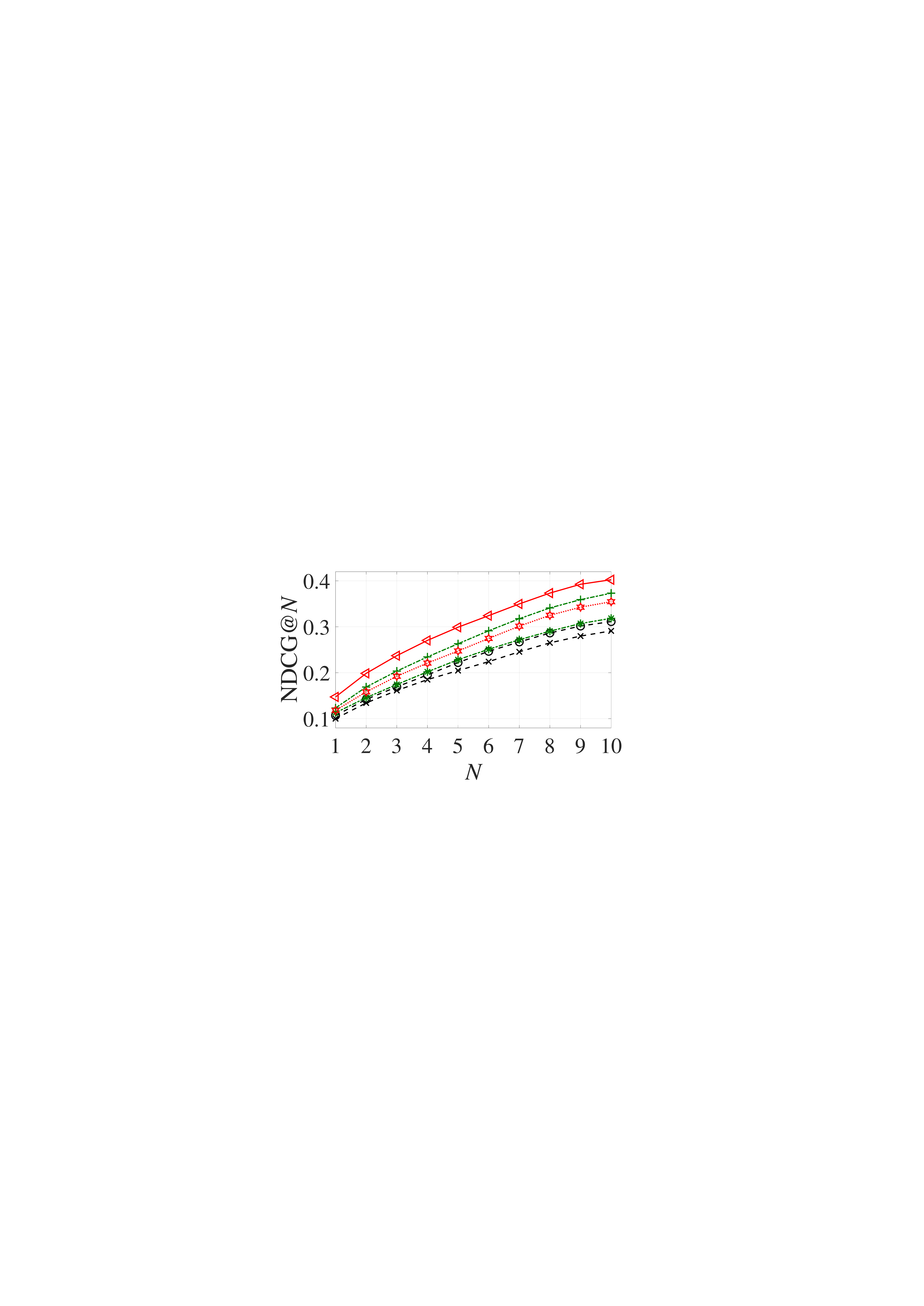}}
	\caption{The result of Top-$N$ recommendation for Task 1 ($k=8$)}
	\label{TopNRecommendation}
\end{figure*}
\begin{table}[t]
	\caption{The experimental results of GA-MTCDR-P ($k=64$) for Task 4 on different sparsity degrees of sub-datasets (density = $1$ - sparsity)} \label{ExperimentalResults_Sparsity}
	\begin{center}
		\resizebox{\columnwidth}{!}{
		\begin{tabular}{c@{    }|@{    }c@{    }|@{    }c@{    }|@{    }c@{    }|@{    }c}
			\hline
			\textbf{Datasets}& \multicolumn{4}{c}{DoubanBook}\\
			\hline
			\textbf{Versions}   & v1 & v2 & v3 & v4 \\
			\hline
			\textbf{\#Users} & 1,413 & 1,050 & 583 & 276\\
			\hline
			\textbf{\#Items} & 6,777 & 6,777    & 6,775 & 6,739\\
			\hline
			\textbf{\#Interactions} & 93,165 & 88,053  & 72,767& 51,581\\
			\hline
			\textbf{Density}& 0.97\% & 1.24\%  & 1.84\% &2.77\%\\
			\hline
			\textbf{HR@$10$~~NDCG@$10$}& .4746 ~ .2928 & .4827 ~ .3021& .4911 ~ .3097 &.4860 ~ .3077 \\			
			\hline
			\hline
			\textbf{Datasets}& \multicolumn{4}{c}{DoubanMusic}\\
			\hline
			\textbf{Versions}   & v1 & v2 & v3 & v4  \\
			\hline
			\textbf{\#Users} & 909 & 634 & 356 & 185\\
			\hline
			\textbf{\#Items} & 5,567 & 5,567 & 5,557 & 5,543\\
			\hline
			\textbf{\#Interactions} & 66,996 & 63,147 & 54,390 & 42,379 \\
			\hline
			\textbf{Density}& 1.32\% & 1.79\%  & 2.75\% &4.13\%\\
			\hline
			\textbf{HR@$10$~~NDCG@$10$}& .4361 ~ .2502 & .4457 ~.2590& .4505 ~ .2617 &.4448 ~ .2581 \\			
			\hline
			\hline
			\textbf{Datasets}& \multicolumn{4}{c}{DoubanMovie}\\
			\hline
			\textbf{Versions}   & v1 & v2 & v3 & v4  \\
			\hline
			\textbf{\#Users} & 2,589 & 2,514 & 2,337 & 2,060\\
			\hline
			\textbf{\#Items} & 9,555 & 9,555    & 9,555 & 9,555\\
			\hline
			\textbf{\#Interactions} & 1,132,973 & 1,131,899  & 1,125,831& 1,104,814\\
			\hline
			\textbf{Density}& 4.58\% & 4.71\%  & 5.04\% &5.61\%\\
			\hline
			\textbf{HR@$10$~~NDCG@$10$}& .6822 ~ .4234 & .6836 ~ .4246& .6910 ~ .4345 &.6815 ~ .4232 \\	
			
			\hline
			\hline
			\textbf{\#Common Users}   & 727 & 444 & 198 & 58 \\
			\hline
		\end{tabular}
	}
    \end{center}
\end{table}
\subsubsection{Result 1: Performance Comparison (for Q1)}
To answer \textbf{Q1}, we compare the performance of our GA-DTCDR-P with those of the seven baseline models. Note that for the SDR baselines, we train them in each domain and then report their performance in each domain; for the single-target CDR baselines, we train them in both domains and then only report their performance on the sparser domain; and for the dual-target CDR models, we train them in both domains and then report their performance in each domain.

Table \ref{ExperimentalResults} shows the experimental results in terms of HR@10 and NDCG@10 with different $k$ embedding dimensions for Tasks 1, 2, and 3, respectively. As indicated in Table \ref{ExperimentalResults}, our GA-DTCDR-P outperforms all the SDR, single-target CDR, and dual-target CDR baselines by an average improvement of 9.04\%. In particular, our GA-DTCDR-P improves the best-performing baselines (with results marked by * in Table \ref{ExperimentalResults}) by an average of 10.85\% for Task 1, an average of 11.21\% for Task 2, and an average of 5.06\% for Task 3. This is because our GA-DTCDR-P effectively leverages the richness and diversity of the information in both domains, and intelligently and effectively combines the embeddings of common users.

Tables \ref{ExperimentalResults_Task4} and \ref{ExperimentalResults_Task5} show the experimental results of our GA-MCDR-P and GA-CDR+CSR-P with different embedding dimensions $k$ for Tasks 4 and 5, respectively. Compared with the results of the seven baselines in Table \ref{ExperimentalResults}, our GA-MTCDR-P and GA-CDR+CSR-P can improve the best-performing baselines by an average of 9.21\% (the general improvement of our GA-DTCDR-P is 9.04\%). This means that, in general, the recommendation performance in all datasets improves as more datasets join in, and hence avoiding negative transfer to some extent.

\subsubsection{Result 2: Ablation Study (for Q2)}
To answer \textbf{Q2}, we implement a variant of our preliminary GA-DTCDR, i.e., GA-DTCDR\_Average, by replacing the element-wise attention with average-pooling, which can demonstrate the detailed contribution of the element-wise attention in our GA models. Average-pooling is the combination strategy used by the existing dual-target CDR approaches \cite{zhu2019dtcdr}, which gives the weight equally, i.e., $0.5$, to the embeddings of common users learned from dual domains. Additionally, to demonstrate the detailed contribution of our proposed personalized training strategy, we also compare the performance of GA-DTCDR with that of GA-DTCDR-P in this section.

On the one hand, as we can see from Table \ref{ExperimentalResults}, with the element-wise attention, our preliminary GA-DTCDR improves GA-DTCDR\_Average by an average of 6.76\%. This means that element-wise attention plays a very important role in our GA-DTCDR and the existing fixed combination strategies can hardly achieve an effective embedding optimization in each target dataset. 

On the other hand, compared with our preliminary GA-DTCDR, our GA-DTCDR-P achieves an average improvement of 0.54\% (according to the results in Table \ref{ExperimentalResults}). This result indicates that our personalized training strategy can further improve the recommendation accuracy of the baselines and our DTCDR models and can alleviate negative transfer.

\subsubsection{Result 3: Impact of Embedding Dimension $k$ (for Q3)}

To answer \textbf{Q3}, we analyze the effect of $k$ on the performance of our preliminary GA-DTCDR, GA-DTCDR-P, GA-MTCDR-P, and GA-CDR+CSR-P, as depicted in Tables \ref{ExperimentalResults}, \ref{ExperimentalResults_Task4}, and \ref{ExperimentalResults_Task5}. In general, in terms of HR@$10$ and NDCG@$10$, the recommendation accuracy of our GA models increases with $k$ because a larger embedding can represent a user/item more accurately. However, considering the structure of the neural network layers in \textbf{Parameter Setting}, the training time of our GA models also increases with $k$. This is a trade-off. Therefore, considering both aspects, $k = 64$ is ideal in our experiments.

\subsubsection{Result 4: Top-$N$ Recommendation (for Q4)}

To answer \textbf{Q4}, we compare the performance of top-$N$ recommendation in terms of HR@$N$ and NDCG@$N$ where $N$ ranges from $1$ to $10$. In fact, the performance trends of all top-$N$ experiments (for all the tasks with different $k$) are similar. Thus, due to space limitation, we only report the Top-$N$ recommendation results of all the seven baseline models, GA-DTCDR\_Average, and GA-DTCDR for Task 1 ($k=8$). In Fig. \ref{TopNRecommendation}, in both DoubanBook (sparser) and DoubanMovie (richer), the performance of our preliminary GA-DTCDR is consistently better than those of all the seven baselines. On DoubanBook, considering all the Top-$N$ recommendations, our preliminary GA-DTCDR improves the best-performing baselines in different experimental cases by an average of 1.74\% for HR@$N$, and by an average of 5.83\% for NDCG@$N$, while on DoubanMovie, our preliminary GA-DTCDR improves the best-performing baselines in different experimental cases by an average of 8.13\% for HR@$N$, and by an average of 7.55\% for NDCG@$N$.

\subsubsection{Result 5: Impact of Sparsity and Overlap Scale (for Q5)}
To answer \textbf{Q5}, we extract different sparsity degrees of sub-datasets to analyze the effect of sparsity and overlap scale on the performance of our models. Due to space limitations, we only report the experimental results of GA-MTCDR-P ($k=64$) for Task 4 on different sub-datasets in Table \ref{ExperimentalResults_Sparsity}. From it, we find that, in general, the recommendation performance increases with densities of sub-datasets. However, the recommendation results in Dataset Version v3 are better than those in Dataset Version v4. This is because the number of common users, i.e., overlap scale, significantly decreases with the increase of density. Therefore, according to the experimental results, in general, the recommendation performance of our GA-MTCDR-P increases with density (i.e., decreases with sparsity) and overlap scale.

\section{Conclusion and Future Work}
In this paper, we have proposed a unified framework, called GA (based on \textbf{G}raph embedding and \textbf{A}ttention techniques), for all dual-target CDR (GA-DTCDR-P), multi-target CDR (GA-MTCDR-P), and CDR+CSR (GA-CDR+CSR-P) scenarios. In our GA framework, the element-wise attention mechanism and the personalized training strategy effectively improve the recommendation accuracy in all datasets and avoid negative transfer to some extent. Also, we have conducted extensive experiments to demonstrate the superior performance of our proposed GA models. In the future, we plan to take more training strategies and further alleviate negative transfer.

\appendices


\ifCLASSOPTIONcompsoc
\else
\fi

\ifCLASSOPTIONcaptionsoff
  \newpage
\fi



\bibliographystyle{IEEEtran}
\bibliography{TKDE2021}

\begin{thebibliography}{10}
\providecommand{\url}[1]{#1}
\csname url@samestyle\endcsname
\providecommand{\newblock}{\relax}
\providecommand{\bibinfo}[2]{#2}
\providecommand{\BIBentrySTDinterwordspacing}{\spaceskip=0pt\relax}
\providecommand{\BIBentryALTinterwordstretchfactor}{4}
\providecommand{\BIBentryALTinterwordspacing}{\spaceskip=\fontdimen2\font plus
\BIBentryALTinterwordstretchfactor\fontdimen3\font minus
  \fontdimen4\font\relax}
\providecommand{\BIBforeignlanguage}[2]{{%
\expandafter\ifx\csname l@#1\endcsname\relax
\typeout{** WARNING: IEEEtran.bst: No hyphenation pattern has been}%
\typeout{** loaded for the language `#1'. Using the pattern for}%
\typeout{** the default language instead.}%
\else
\language=\csname l@#1\endcsname
\fi
#2}}
\providecommand{\BIBdecl}{\relax}
\BIBdecl

\bibitem{berkovsky2007cross}
S.~Berkovsky, T.~Kuflik, and F.~Ricci, ``Cross-domain mediation in
  collaborative filtering,'' in \emph{International Conference on User
  Modeling}.\hskip 1em plus 0.5em minus 0.4em\relax Springer, 2007, pp.
  355--359.

\bibitem{zhao2013active}
L.~Zhao, S.~J. Pan, E.~W. Xiang, E.~Zhong, Z.~Lu, and Q.~Yang, ``Active
  transfer learning for cross-system recommendation,'' in \emph{Twenty-Seventh
  AAAI Conference on Artificial Intelligence}, 2013.

\bibitem{zhu2018deep}
F.~Zhu, Y.~Wang, C.~Chen, G.~Liu, M.~A. Orgun, and J.~Wu, ``A deep framework
  for cross-domain and cross-system recommendations,'' in \emph{IJCAI
  International Joint Conference on Artificial Intelligence}, 2018, pp.
  3711--3717.

\bibitem{tang2012cross}
J.~Tang, S.~Wu, J.~Sun, and H.~Su, ``Cross-domain collaboration
  recommendation,'' in \emph{Proceedings of the 18th ACM SIGKDD international
  conference on Knowledge discovery and data mining}, 2012, pp. 1285--1293.

\bibitem{fu2019deeply}
W.~Fu, Z.~Peng, S.~Wang, Y.~Xu, and J.~Li, ``Deeply fusing reviews and contents
  for cold start users in cross-domain recommendation systems,'' in
  \emph{Proceedings of the AAAI Conference on Artificial Intelligence},
  vol.~33, 2019, pp. 94--101.

\bibitem{shapira2013facebook}
B.~Shapira, L.~Rokach, and S.~Freilikhman, ``Facebook single and cross domain
  data for recommendation systems,'' \emph{User Modeling and User-Adapted
  Interaction}, vol.~23, no. 2-3, pp. 211--247, 2013.

\bibitem{jiang2012social}
M.~Jiang, P.~Cui, F.~Wang, Q.~Yang, W.~Zhu, and S.~Yang, ``Social
  recommendation across multiple relational domains,'' in \emph{Proceedings of
  the 21st ACM international conference on Information and knowledge
  management}, 2012, pp. 1422--1431.

\bibitem{jiang2015social}
M.~Jiang, P.~Cui, X.~Chen, F.~Wang, W.~Zhu, and S.~Yang, ``Social
  recommendation with cross-domain transferable knowledge,'' \emph{IEEE
  transactions on knowledge and data engineering}, vol.~27, no.~11, pp.
  3084--3097, 2015.

\bibitem{tan2014cross}
S.~Tan, J.~Bu, X.~Qin, C.~Chen, and D.~Cai, ``Cross domain recommendation based
  on multi-type media fusion,'' \emph{Neurocomputing}, vol. 127, pp. 124--134,
  2014.

\bibitem{fernandez2014exploiting}
I.~Fern{\'a}ndez-Tob{\'\i}as and I.~Cantador, ``Exploiting social tags in
  matrix factorization models for cross-domain collaborative filtering.'' in
  \emph{CBRecSys@ RecSys}, 2014, pp. 34--41.

\bibitem{zhang2017cross}
Q.~Zhang, D.~Wu, J.~Lu, F.~Liu, and G.~Zhang, ``A cross-domain recommender
  system with consistent information transfer,'' \emph{Decision Support
  Systems}, vol. 104, pp. 49--63, 2017.

\bibitem{liu2017mcs}
G.~Liu, Y.~Liu, K.~Zheng, A.~Liu, Z.~Li, Y.~Wang, and X.~Zhou, ``Mcs-gpm:
  Multi-constrained simulation based graph pattern matching in contextual
  social graphs,'' \emph{IEEE Transactions on Knowledge and Data Engineering},
  vol.~30, no.~6, pp. 1050--1064, 2017.

\bibitem{zhang2018cross2}
Q.~Zhang, D.~Wu, J.~Lu, and G.~Zhang, ``Cross-domain recommendation with
  probabilistic knowledge transfer,'' in \emph{International Conference on
  Neural Information Processing}.\hskip 1em plus 0.5em minus 0.4em\relax
  Springer, 2018, pp. 208--219.

\bibitem{hu2019transfer}
G.~Hu, Y.~Zhang, and Q.~Yang, ``Transfer meets hybrid: A synthetic approach for
  cross-domain collaborative filtering with text,'' in \emph{The World Wide Web
  Conference}, 2019, pp. 2822--2829.

\bibitem{manotumruksa2019cross}
J.~Manotumruksa, D.~Rafailidis, C.~Macdonald, and I.~Ounis, ``On cross-domain
  transfer in venue recommendation,'' in \emph{European Conference on
  Information Retrieval}.\hskip 1em plus 0.5em minus 0.4em\relax Springer,
  2019, pp. 443--456.

\bibitem{huang2019lscd}
L.~Huang, Z.-L. Zhao, C.-D. Wang, D.~Huang, and H.-Y. Chao, ``Lscd: Low-rank
  and sparse cross-domain recommendation,'' \emph{Neurocomputing}, vol. 366,
  pp. 86--96, 2019.

\bibitem{li2019cross}
L.~Li, Q.~Do, and W.~Liu, ``Cross-domain recommendation via coupled
  factorization machines,'' in \emph{Proceedings of the AAAI Conference on
  Artificial Intelligence}, vol.~33, 2019, pp. 9965--9966.

\bibitem{zhao2020catn}
C.~Zhao, C.~Li, R.~Xiao, H.~Deng, and A.~Sun, ``Catn: Cross-domain
  recommendation for cold-start users via aspect transfer network,''
  \emph{arXiv preprint arXiv:2005.10549}, 2020.

\bibitem{kang2020deep}
Y.~Kang, S.~Gai, F.~Zhao, D.~Wang, and A.~Tang, ``Deep transfer collaborative
  filtering with geometric structure preservation for cross-domain
  recommendation,'' in \emph{2020 International Joint Conference on Neural
  Networks (IJCNN)}.\hskip 1em plus 0.5em minus 0.4em\relax IEEE, 2020, pp.
  1--8.

\bibitem{rendle2009bpr}
S.~Rendle, C.~Freudenthaler, Z.~Gantner, and L.~Schmidt-Thieme, ``Bpr: Bayesian
  personalized ranking from implicit feedback,'' in \emph{UAI}, 2009, pp.
  452--461.

\bibitem{he2017neural}
X.~He, L.~Liao, H.~Zhang, L.~Nie, X.~Hu, and T.-S. Chua, ``Neural collaborative
  filtering,'' in \emph{Proceedings of the 26th international conference on
  world wide web}, 2017, pp. 173--182.

\bibitem{xue2017deep}
H.-J. Xue, X.~Dai, J.~Zhang, S.~Huang, and J.~Chen, ``Deep matrix factorization
  models for recommender systems,'' in \emph{Proceedings of the 26th
  international joint conference on artificial intelligence}, 2017, pp.
  3203--3209.

\bibitem{zhang2016multi}
Z.~Zhang, X.~Jin, L.~Li, G.~Ding, and Q.~Yang, ``Multi-domain active learning
  for recommendation,'' in \emph{Thirtieth AAAI Conference on Artificial
  Intelligence}, 2016, pp. 2358--2364.

\bibitem{man2017cross}
T.~Man, H.~Shen, X.~Jin, and X.~Cheng, ``Cross-domain recommendation: An
  embedding and mapping approach,'' in \emph{Proceedings of the Twenty-Sixth
  International Joint Conference on Artificial Intelligence}, 2017, pp.
  2464--2470.

\bibitem{zhu2019dtcdr}
F.~Zhu, C.~Chen, Y.~Wang, G.~Liu, and X.~Zheng, ``Dtcdr: A framework for
  dual-target cross-domain recommendation,'' in \emph{Proceedings of the 28th
  ACM International Conference on Information and Knowledge Management}.\hskip
  1em plus 0.5em minus 0.4em\relax ACM, 2019, pp. 1533--1542.

\bibitem{li2019ddtcdr}
P.~Li and A.~Tuzhilin, ``Ddtcdr: Deep dual transfer cross domain
  recommendation,'' \emph{arXiv preprint arXiv:1910.05189}, 2019.

\bibitem{zhu2020graphical}
F.~Zhu, Y.~Wang, C.~Chen, G.~Liu, and X.~Zheng, ``A graphical and attentional
  framework for dual-target cross-domain recommendation,'' in \emph{29th
  International Joint Conference on Artificial Intelligence}, 2020, pp.
  3001--3008.

\bibitem{li2021dual}
P.~Li and A.~Tuzhilin, ``Dual metric learning for effective and efficient
  cross-domain recommendations,'' \emph{IEEE Transactions on Knowledge and Data
  Engineering}, 2021.

\bibitem{pan2009survey}
S.~J. Pan and Q.~Yang, ``A survey on transfer learning,'' \emph{TKDE}, vol.~22,
  no.~10, pp. 1345--1359, 2009.

\bibitem{zhang2012multi}
Y.~Zhang, B.~Cao, and D.-Y. Yeung, ``Multi-domain collaborative filtering,''
  \emph{arXiv preprint arXiv:1203.3535}, 2012.

\bibitem{moreno2012talmud}
O.~Moreno, B.~Shapira, L.~Rokach, and G.~Shani, ``Talmud: transfer learning for
  multiple domains,'' in \emph{Proceedings of the 21st ACM international
  conference on Information and knowledge management}, 2012, pp. 425--434.

\bibitem{pan2013transfer}
W.~Pan and Q.~Yang, ``Transfer learning in heterogeneous collaborative
  filtering domains,'' \emph{Artificial intelligence}, vol. 197, pp. 39--55,
  2013.

\bibitem{liu2020cross}
M.~Liu, J.~Li, G.~Li, and P.~Pan, ``Cross domain recommendation via
  bi-directional transfer graph collaborative filtering networks,'' in
  \emph{Proceedings of the 29th ACM International Conference on Information \&
  Knowledge Management}, 2020, pp. 885--894.

\bibitem{shi2011tags}
Y.~Shi, M.~Larson, and A.~Hanjalic, ``Tags as bridges between domains:
  Improving recommendation with tag-induced cross-domain collaborative
  filtering,'' in \emph{International Conference on User Modeling, Adaptation,
  and Personalization}.\hskip 1em plus 0.5em minus 0.4em\relax Springer, 2011,
  pp. 305--316.

\bibitem{wang2020tag}
J.~Wang and J.~Lv, ``Tag-informed collaborative topic modeling for cross domain
  recommendations,'' \emph{Knowledge-Based Systems}, vol. 203, pp. 106--119,
  2020.

\bibitem{jiang2016little}
M.~Jiang, P.~Cui, N.~J. Yuan, X.~Xie, and S.~Yang, ``Little is much: Bridging
  cross-platform behaviors through overlapped crowds,'' in \emph{Proceedings of
  the AAAI Conference on Artificial Intelligence}, vol.~30, no.~1, 2016.

\bibitem{zhang2019cross}
Q.~Zhang, P.~Hao, J.~Lu, and G.~Zhang, ``Cross-domain recommendation with
  semantic correlation in tagging systems,'' in \emph{2019 International Joint
  Conference on Neural Networks (IJCNN)}.\hskip 1em plus 0.5em minus
  0.4em\relax IEEE, 2019, pp. 1--8.

\bibitem{sahebi2014content}
S.~Sahebi and T.~Walker, ``Content-based cross-domain recommendations using
  segmented models.'' in \emph{CBRecSys@ RecSys}, 2014, pp. 57--64.

\bibitem{kanagawa2019cross}
H.~Kanagawa, H.~Kobayashi, N.~Shimizu, Y.~Tagami, and T.~Suzuki, ``Cross-domain
  recommendation via deep domain adaptation,'' in \emph{European Conference on
  Information Retrieval}.\hskip 1em plus 0.5em minus 0.4em\relax Springer,
  2019, pp. 20--29.

\bibitem{lu2018like}
Y.~Lu, R.~Dong, and B.~Smyth, ``Why i like it: multi-task learning for
  recommendation and explanation,'' in \emph{Proceedings of the 12th ACM
  Conference on Recommender Systems}, 2018, pp. 4--12.

\bibitem{hu2018conet}
G.~Hu, Y.~Zhang, and Q.~Yang, ``Conet: Collaborative cross networks for
  cross-domain recommendation,'' in \emph{Proceedings of the 27th ACM
  International Conference on Information and Knowledge Management}, 2018, pp.
  667--676.

\bibitem{he2018robust}
M.~He, J.~Zhang, P.~Yang, and K.~Yao, ``Robust transfer learning for
  cross-domain collaborative filtering using multiple rating patterns
  approximation,'' in \emph{Proceedings of the Eleventh ACM International
  Conference on Web Search and Data Mining}, 2018, pp. 225--233.

\bibitem{wang2019solving}
Y.~Wang, C.~Feng, C.~Guo, Y.~Chu, and J.-N. Hwang, ``Solving the sparsity
  problem in recommendations via cross-domain item embedding based on
  co-clustering,'' in \emph{Proceedings of the Twelfth ACM International
  Conference on Web Search and Data Mining}, 2019, pp. 717--725.

\bibitem{liu2018transferable}
B.~Liu, Y.~Wei, Y.~Zhang, Z.~Yan, and Q.~Yang, ``Transferable contextual bandit
  for cross-domain recommendation,'' in \emph{Thirty-Second AAAI Conference on
  Artificial Intelligence}, 2018, pp. 3619--3626.

\bibitem{liu2020exploiting}
J.~Liu, P.~Zhao, F.~Zhuang, Y.~Liu, V.~S. Sheng, J.~Xu, X.~Zhou, and H.~Xiong,
  ``Exploiting aesthetic preference in deep cross networks for cross-domain
  recommendation,'' in \emph{Proceedings of The Web Conference 2020}, 2020, pp.
  2768--2774.

\bibitem{sopchoke2018explainable}
S.~Sopchoke, K.-i. Fukui, and M.~Numao, ``Explainable cross-domain
  recommendations through relational learning,'' in \emph{Thirty-Second AAAI
  Conference on Artificial Intelligence}, 2018.

\bibitem{kang2019semi}
S.~Kang, J.~Hwang, D.~Lee, and H.~Yu, ``Semi-supervised learning for
  cross-domain recommendation to cold-start users,'' in \emph{Proceedings of
  the 28th ACM International Conference on Information and Knowledge
  Management}, 2019, pp. 1563--1572.

\bibitem{yuan2019darec}
F.~Yuan, L.~Yao, and B.~Benatallah, ``Darec: Deep domain adaptation for
  cross-domain recommendation via transferring rating patterns,'' \emph{arXiv
  preprint arXiv:1905.10760}, 2019.

\bibitem{zhang2020cross}
Q.~Zhang, J.~Lu, and G.~Zhang, ``Cross-domain recommendation with multiple
  sources,'' in \emph{2020 International Joint Conference on Neural Networks
  (IJCNN)}.\hskip 1em plus 0.5em minus 0.4em\relax IEEE, 2020, pp. 1--7.

\bibitem{zhou2018graph}
J.~Zhou, G.~Cui, Z.~Zhang, C.~Yang, Z.~Liu, and M.~Sun, ``Graph neural
  networks: A review of methods and applications,'' \emph{arXiv preprint
  arXiv:1812.08434}, 2018.

\bibitem{kruskal1978multidimensional}
J.~B. Kruskal, \emph{Multidimensional scaling}.\hskip 1em plus 0.5em minus
  0.4em\relax Sage, 1978, no.~11.

\bibitem{wold1987principal}
S.~Wold, K.~Esbensen, and P.~Geladi, ``Principal component analysis,''
  \emph{Chemometrics and intelligent laboratory systems}, vol.~2, no. 1-3, pp.
  37--52, 1987.

\bibitem{yan2005graph}
S.~Yan, D.~Xu, B.~Zhang, and H.-J. Zhang, ``Graph embedding: A general
  framework for dimensionality reduction,'' in \emph{2005 IEEE Computer Society
  Conference on Computer Vision and Pattern Recognition (CVPR'05)},
  vol.~2.\hskip 1em plus 0.5em minus 0.4em\relax IEEE, 2005, pp. 830--837.

\bibitem{perozzi2014deepwalk}
B.~Perozzi, R.~Al-Rfou, and S.~Skiena, ``Deepwalk: Online learning of social
  representations,'' in \emph{Proceedings of the 20th ACM SIGKDD international
  conference on Knowledge discovery and data mining}, 2014, pp. 701--710.

\bibitem{tang2015line}
J.~Tang, M.~Qu, M.~Wang, M.~Zhang, J.~Yan, and Q.~Mei, ``Line: Large-scale
  information network embedding,'' in \emph{Proceedings of the 24th
  international conference on world wide web}, 2015, pp. 1067--1077.

\bibitem{grover2016node2vec}
A.~Grover and J.~Leskovec, ``node2vec: Scalable feature learning for
  networks,'' in \emph{Proceedings of the 22nd ACM SIGKDD international
  conference on Knowledge discovery and data mining}, 2016, pp. 855--864.

\bibitem{zhao2019intentgc}
J.~Zhao, Z.~Zhou, Z.~Guan, W.~Zhao, W.~Ning, G.~Qiu, and X.~He, ``Intentgc: a
  scalable graph convolution framework fusing heterogeneous information for
  recommendation,'' in \emph{Proceedings of the 25th ACM SIGKDD International
  Conference on Knowledge Discovery \& Data Mining}, 2019, pp. 2347--2357.

\bibitem{yang2020interpretable}
Y.~Yang, Z.~Guan, J.~Li, J.~Huang, and W.~Zhao, ``Interpretable and efficient
  heterogeneous graph convolutional network,'' \emph{arXiv preprint
  arXiv:2005.13183}, 2020.

\bibitem{bahdanau2014neural}
D.~Bahdanau, K.~Cho, and Y.~Bengio, ``Neural machine translation by jointly
  learning to align and translate,'' \emph{arXiv preprint arXiv:1409.0473},
  2014.

\bibitem{chen2017attentive}
J.~Chen, H.~Zhang, X.~He, L.~Nie, W.~Liu, and T.-S. Chua, ``Attentive
  collaborative filtering: Multimedia recommendation with item-and
  component-level attention,'' in \emph{Proceedings of the 40th International
  ACM SIGIR conference on Research and Development in Information Retrieval},
  2017, pp. 335--344.

\bibitem{hu2018leveraging}
B.~Hu, C.~Shi, W.~X. Zhao, and P.~S. Yu, ``Leveraging meta-path based context
  for top-n recommendation with a neural co-attention model,'' in
  \emph{Proceedings of the 24th ACM SIGKDD International Conference on
  Knowledge Discovery \& Data Mining}, 2018, pp. 1531--1540.

\bibitem{tay2018multi}
Y.~Tay, A.~T. Luu, and S.~C. Hui, ``Multi-pointer co-attention networks for
  recommendation,'' in \emph{Proceedings of the 24th ACM SIGKDD International
  Conference on Knowledge Discovery \& Data Mining}, 2018, pp. 2309--2318.

\bibitem{wang2019kgat}
X.~Wang, X.~He, Y.~Cao, M.~Liu, and T.-S. Chua, ``Kgat: Knowledge graph
  attention network for recommendation,'' in \emph{Proceedings of the 25th ACM
  SIGKDD International Conference on Knowledge Discovery \& Data Mining}, 2019,
  pp. 950--958.

\bibitem{zhao2017unified}
L.~Zhao, S.~J. Pan, and Q.~Yang, ``A unified framework of active transfer
  learning for cross-system recommendation,'' \emph{Artificial Intelligence},
  vol. 245, pp. 38--55, 2017.

\bibitem{le2014distributed}
Q.~Le and T.~Mikolov, ``Distributed representations of sentences and
  documents,'' in \emph{ICML}, 2014, pp. 1188--1196.

\bibitem{manning2014}
C.~D. Manning, M.~Surdeanu, J.~Bauer, J.~Finkel, S.~J. Bethard, and
  D.~McClosky, ``The {Stanford} {CoreNLP} toolkit,'' in \emph{ACL System
  Demonstrations}, 2014, pp. 55--60.

\bibitem{xin2015cross}
X.~Xin, Z.~Liu, C.-Y. Lin, H.~Huang, X.~Wei, and P.~Guo, ``Cross-domain
  collaborative filtering with review text,'' in \emph{Twenty-Fourth
  International Joint Conference on Artificial Intelligence}, 2015, pp.
  1827--1834.

\bibitem{yang2020federated}
L.~Yang, B.~Tan, V.~W. Zheng, K.~Chen, and Q.~Yang, ``Federated recommendation
  systems,'' in \emph{Federated Learning}.\hskip 1em plus 0.5em minus
  0.4em\relax Springer, 2020, pp. 225--239.

\bibitem{huang2020personalized}
Y.~Huang, L.~Chu, Z.~Zhou, L.~Wang, J.~Liu, J.~Pei, and Y.~Zhang,
  ``Personalized federated learning: An attentive collaboration approach,''
  \emph{arXiv preprint arXiv:2007.03797}, 2020.

\bibitem{harper2016movielens}
F.~M. Harper and J.~A. Konstan, ``The movielens datasets: History and
  context,'' \emph{Acm transactions on interactive intelligent systems (TIIS)},
  vol.~5, no.~4, p.~19, 2016.

\bibitem{kingma2014adam}
D.~P. Kingma and J.~Ba, ``Adam: A method for stochastic optimization,''
  \emph{arXiv preprint arXiv:1412.6980}, 2014.

\end{thebibliography}
%



%

\begin{IEEEbiography}[{\includegraphics[width=1in,height=1.25in,clip,keepaspectratio]{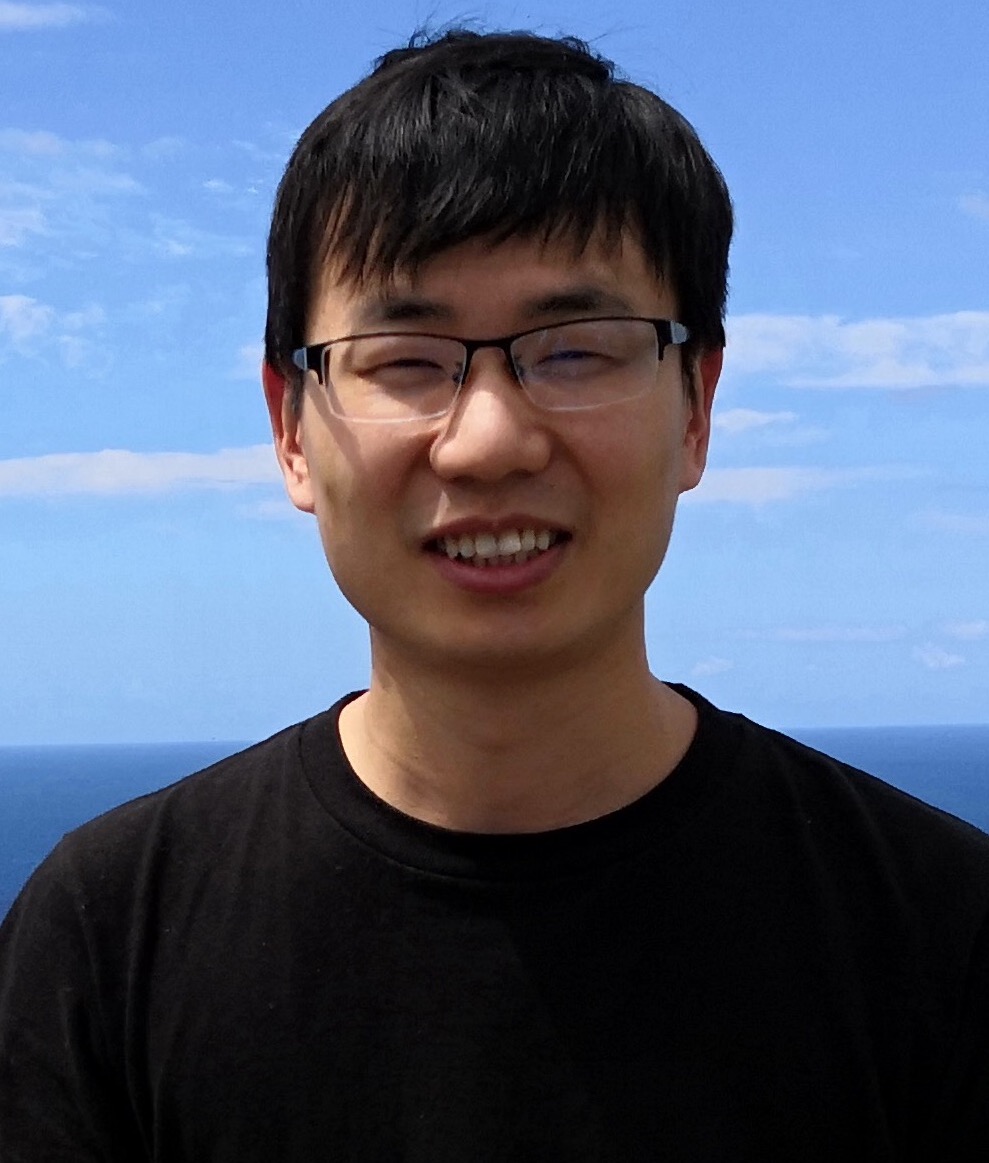}}]{Feng Zhu}
received his PhD degree in Computer Science from Macquarie University, Australia, in 2020. He is currently a senior algorithm engineer at Ant Group, Hangzhou, P.R. China. His research interests include social computing, cross-domain recommender systems, and machine learning. He has published three regular papers and a survey paper (accepted) in the area of cross-domain recommendation in peer reviewed conferences.
\end{IEEEbiography}

\begin{IEEEbiography}[{\includegraphics[width=1in,height=1.25in,clip,keepaspectratio]{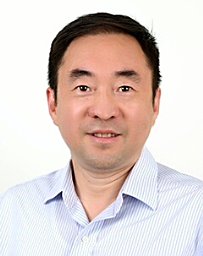}}]{Yan Wang}
received the BEng, MEng, and the DEng degrees in computer science and technology from Harbin Institute of Technology (HIT), P. R. China, in 1988, 1991, and 1996, respectively. He is currently a Professor in the Department of Computing, Macquarie University, Sydney, Australia. His research interests include trust computing, recommender systems, social computing, and service computing. He is a senior member of the IEEE.
\end{IEEEbiography}

\begin{IEEEbiography}[{\includegraphics[width=1in,height=1.25in,clip,keepaspectratio]{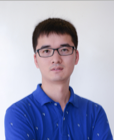}}]{Jun Zhou}
is currently a Senior Staff Engineer at Ant Group. His research mainly focuses on machine learning and data mining. He has participated in the development of several distributed systems and machine learning platforms in Alibaba and Ant Financial, such as Apsaras (Distributed Operating System) and KunPeng (Parameter Server). He has published more than 40 papers in top-tier machine learning and data mining conferences.
\end{IEEEbiography}

\begin{IEEEbiography}[{\includegraphics[width=1in,height=1.25in,clip,keepaspectratio]{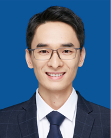}}]{Chaochao Chen}
obtained his PhD degree in computer science from Zhejiang University, China, in 2016, and he was a visiting scholar in University of Illinois at Urbana-Champaign, during 2014-2015. He is currently a Staff Algorithm Engineer at Ant Group. His research mainly focuses on recommender system, privacy preserving machine learning, transfer learning, graph representation, and distributed machine learning. He has published more than 40 papers in peer reviewed journals and conferences.
\end{IEEEbiography}

\begin{IEEEbiography}[{\includegraphics[width=1in,height=1.25in,clip,keepaspectratio]{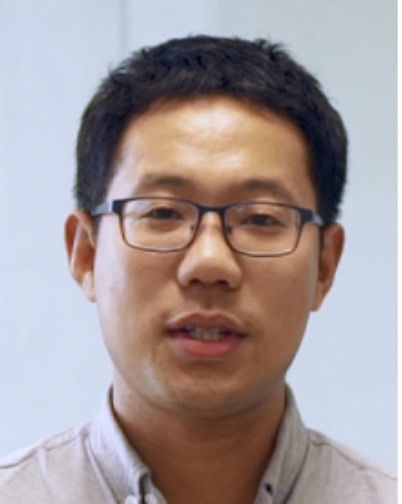}}]{Longfei Li}
 is currently a Staff Algorithm Engineer at Ant Group. His research mainly focuses on machine learning, AutoML, and Causal Infernece. He has published more than 10 papers in top-tier machine learning and data mining conferences.
\end{IEEEbiography}

\begin{IEEEbiography}[{\includegraphics[width=1in,height=1.25in,clip,keepaspectratio]{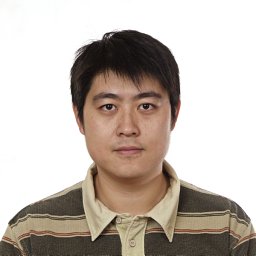}}]{Guanfeng Liu}
is currently a Lecturer in the Department of Computing, Macquarie University, Sydney, Australia. He received his PhD degree in Computer Science from Macquarie University in 2013. His research interests include graph database, trust computing, and social computing.
\end{IEEEbiography}





\end{document}